\documentclass[twocolumn,times]{aastex62}
\usepackage{graphicx}
\usepackage[flushleft]{threeparttable}
\usepackage{blindtext}
\usepackage{amsmath}
\usepackage{mathtools}
\usepackage{multirow}
\usepackage{comment}
\turnoffeditone

\begin{document}
\shortauthors{Jung et al.}
\def\nar{New Astron.}
\def\na{New Astron.}
\title{\large \textbf{CLEAR: Boosted Ly$\alpha$ Transmission of the Intergalactic Medium in UV bright Galaxies}}

\correspondingauthor{Intae Jung}
\email{intae.jung@nasa.gov}

\author[0000-0003-1187-4240]{Intae Jung}
\affil{Astrophysics Science Division, Goddard Space Flight Center, Greenbelt, MD 20771, USA}
\affil{Department of Physics, The Catholic University of America, Washington, DC 20064, USA }
\affil{Center for Research and Exploration in Space Science and Technology, NASA/GSFC, Greenbelt, MD 20771}

\author[0000-0001-7503-8482]{Casey Papovich}
\affil{Department of Physics and Astronomy, Texas A\&M University, College Station, TX, 77843-4242 USA}
\affil{George P.\ and Cynthia Woods Mitchell Institute for Fundamental Physics and Astronomy, Texas A\&M University, College Station, TX, 77843-4242 USA}

\author[0000-0001-8519-1130]{Steven L. Finkelstein}
\affil{Department of Astronomy, The University of Texas at Austin, Austin, TX, 78759 USA}

\author[0000-0002-6386-7299]{Raymond C. Simons}
\affil{Space Telescope Science Institute, 3700 San Martin Drive, Baltimore, MD, 21218 US}

\author[0000-0001-8489-2349]{Vicente Estrada-Carpenter}
\affil{Department of Physics and Astronomy, Texas A\&M University, College Station, TX, 77843-4242 USA}
\affil{George P.\ and Cynthia Woods Mitchell Institute for Fundamental Physics and Astronomy, Texas A\&M University, College Station, TX, 77843-4242 USA}

\author[0000-0001-8534-7502]{Bren E. Backhaus}
\affil{Department of Physics, University of Connecticut, Storrs, CT 06269, USA}

\author[0000-0001-7151-009X]{Nikko J. Cleri}
\affil{Department of Physics and Astronomy, Texas A\&M University, College Station, TX, 77843-4242 USA}
\affil{George P.\ and Cynthia Woods Mitchell Institute for Fundamental Physics and Astronomy, Texas A\&M University, College Station, TX, 77843-4242 USA}

\author[0000-0002-0496-1656]{Kristian Finlator}
\affil{Department of Astronomy, New Mexico State University, Las Cruces, New Mexico, USA}
\affil{Cosmic Dawn Center (DAWN) at the Niels Bohr Institute, University of Copenhagen and National Space Institute, Technical University of Denmark}

\author[0000-0002-7831-8751]{Mauro Giavalisco}
\affil{Astronomy Department, University of Massachusetts, Amherst, MA, 01003 USA} 

\author[0000-0001-7673-2257]{Zhiyuan Ji}
\affil{Astronomy Department, University of Massachusetts, Amherst, MA, 01003 USA} 

\author[0000-0002-7547-3385]{Jasleen Matharu}
\affil{Department of Physics and Astronomy, Texas A\&M University, College Station, TX, 77843-4242 USA}
\affil{George P.\ and Cynthia Woods Mitchell Institute for Fundamental Physics and Astronomy, Texas A\&M University, College Station, TX, 77843-4242 USA}

\author[0000-0003-1665-2073]{Ivelina Momcheva}
\affil{Space Telescope Science Institute, 3700 San Martin Drive, Baltimore, MD, 21218 US}

\author[0000-0002-4772-7878]{Amber N. Straughn}
\affil{Astrophysics Science Division, Goddard Space Flight Center, Greenbelt, MD 20771, USA}

\author[0000-0002-1410-0470]{Jonathan R. Trump}
\affil{Department of Physics, University of Connecticut, Storrs, CT 06269, USA}

\begin{abstract}
Reionization is an inhomogeneous process, thought to begin in small ionized bubbles of the intergalactic medium (IGM) around overdense regions of galaxies.  Recent Lyman-alpha (Ly$\alpha$) studies during the epoch of reionization show growing evidence that ionized bubbles formed earlier around brighter galaxies, suggesting higher IGM transmission of Ly$\alpha$ from these galaxies.  We investigate this problem using IR slitless spectroscopy from the Hubble Space Telescope (HST) Wide-Field Camera 3 (WFC3) {G102 grism observations} of 148 galaxies selected via photometric redshifts at $6.0<z<8.2$. These galaxies have spectra extracted from the CANDELS Ly$\alpha$ Emission at Reionization (CLEAR) survey. \deleted{CLEAR includes 12-orbit depth HST/WFC3 G102 grism observations over 12 WFC3 fields within the CANDELS GOODS survey.}We combine the CLEAR data for 275 galaxies with the Keck/DEIMOS+MOSFIRE dataset from the Texas Spectroscopic Search for Ly$\alpha$ Emission at the End of Reionization Survey. We then constrain the Ly$\alpha$ equivalent-width (EW) distribution at $6.0<z<8.2${, which is described by an exponential form, $dN/d\text{EW}\propto\text{exp(-EW)}/W_0$, with the characteristic $e$-folding scale width ($W_0$). We confirm} a significant drop of the Ly$\alpha$ strength {(or $W_0$)} at $z>6$.  Furthermore, we compare the redshift evolution of \replaced{the Ly$\alpha$ EW distribution}{$W_0$} between galaxies at different UV luminosities.  The UV-bright ($M_{\text{UV}}<-21${, or $L_{\text{UV}}>L^{*}$}) galaxies show weaker evolution with a decrease of 0.4 ($\pm$0.2) dex in \deleted{the $e$-folding scale width (}$W_0$\deleted{)} \deleted{of the Ly$\alpha$ EW distribution} at $z>6$ while UV-faint ($M_{\text{UV}}>-21${, or $L_{\text{UV}}<L^{*}$}) galaxies exhibit a significant drop by a factor of 0.7-0.8 ($\pm0.2$) dex in $W_0$ from $z<6$ to $z>6$.  Our results add to the accumulating evidence that UV-bright galaxies exhibit boosted Ly$\alpha$ transmission in the IGM, suggesting that reionization completes sooner in regions proximate to galaxies of higher UV luminosity.
\end{abstract}


\section{Introduction}
Reionization marks the last major phase transition of the universe, when the first stars and galaxies ionized the intergalactic medium (IGM) neutral hydrogen (\ion{H}{1}) gas.  Galaxies in the early universe are inherently coupled with the process of reionization, as the galaxies were likely the primary sources of ionizing photons \citep[e.g.,][]{Robertson2015a, McQuinn2016a, Finkelstein2019a, Dayal2018a} {while the contribution from active galactic nucleus (AGN) activity could also play an important (though subdominant) role at $z\gtrsim 6$
\citep[e.g.,][]{Matsuoka2018a, Kulkarni2019a, Robertson2021a}}.  Thus, observations of this epoch also provide key information for studying the dominant source of the ionizing photons: galaxies in the early universe.  However, constraining the ionizing photon budget during the epoch of reionization (EoR) is still extremely difficult due to current observational limits as well as a poorly constrained ionizing photon escape fraction which depends on galaxy physical conditions \citep[e.g.,][]{Finkelstein2012a, Finkelstein2015a, Finkelstein2019b, Robertson2013a, Robertson2015a, Bouwens2015a, Bouwens2016a, Kimm2019a, Yoo2020a, Ocvirk2021a}.  Understanding the temporal and spatial evolution of reionization by tracing the IGM \ion{H}{1} fraction provides a key constraint on the ionizing emissivity required from galaxies as a function of redshift.

As Lyman-alpha (Ly$\alpha$) photons are resonantly scattered by \ion{H}{1} in the IGM, an analysis of Ly$\alpha$ can be used to trace the existence of \ion{H}{1} gas in the IGM at different points in the history of the universe \citep[e.g.,][]{Miralda1998a, Rhoads2001a, Malhotra2004a, Dijkstra2014a}. This technique uses follow-up spectroscopic observations, targeting high-$z$ candidate galaxies, to measure the strength of Ly$\alpha$ emission from galaxies in the reionization era. Initial studies using Ly$\alpha$ spectroscopy have found an apparent deficit of Ly$\alpha$ emission at $z > 6.5$ \citep[e.g.,][]{Fontana2010a, Pentericci2011a, Finkelstein2013a, Pentericci2014a, Caruana2012a, Caruana2014a, Curtis-Lake2012a, Mallery2012a, Ono2012a, Schenker2012a, Schenker2014a, Treu2012a, Treu2013a, Tilvi2014a, Vanzella2014a, Schmidt2016a, De-Barros2017a, Fuller2020a}, implying an increasing \ion{H}{1} fraction in the IGM from $z\sim6$ $\rightarrow$ 7, although other Ly$\alpha$ systematics with galaxy evolutionary features are a factor in these measurements \citep[e.g.,][]{Finkelstein2012b, Yang2017a, Tang2019a, Trainor2019a, Du2020a, Hassan2021a, Weiss2021a}.

Recent Ly$\alpha$ studies suggest a more complicated picture of reionization.  For instance, \cite{Pentericci2018b} provide  Ly$\alpha$ fraction ($f_\mathrm{Ly\alpha}$) measurements at $z\sim6$ and $z\sim7$ ($f_\mathrm{Ly\alpha}=N_{\text{LAE}}/N_{\text{LBG}}$, where $N_{\text{LAE}}$ is the number of spectroscopically-confirmed Ly$\alpha$-emitting galaxies, and $N_{\text{LBG}}$ is the number of high-$z$ candidate galaxies that were targeted in spectroscopic observations). Their results show a possible flattening or a steady increase in the redshift evolution the Ly$\alpha$ fraction from $z\sim5 \rightarrow 6$ and a relatively smoother evolution from $z\sim6 \rightarrow 7$, compared to previous studies \citep[e.g.,][]{Stark2011a, Tilvi2014a}, implying a more extended ending to reionization {\citep[this finding is also supported by][]{Kulkarni2019a, Fuller2020a}.}  Furthermore, while \cite{Zheng2017a}, \cite{Castellano2018a}, and \cite{Tilvi2020a} report observations of an ionized bubble via Ly$\alpha$ observations at $z\gtrsim7$, non/rare detections of Ly$\alpha$ in \cite{Hoag2019a} and \cite{Mason2019a} suggest a significantly neutral fraction in the IGM at $z\sim7.5$. Specifically, \cite{Hoag2019a} report a high neutral fraction of $88^{+5}_{-10}$\% at $z \sim 7.6$.  More recently, \cite{Jung2020a} analyzed  deep NIR observations in GOODS-N, suggesting a modestly-ionized universe with the inferred IGM neutral fraction of $49^{+19}_{-19}$\% at $z \sim 7.6$, lower than other Ly$\alpha$ studies at the same redshifts.  Additionally, \cite{Hu2021a} disclose a protocluster structure at $z\sim6.9$ which consists of 16 spectroscopically-confirmed Ly$\alpha$ emitting galaxies.  One way to reconcile these apparently contrasting recent findings is if reionization is complex and inhomogenous, and/or if there are large spatial and temporal variations in the history of reionization.

Taken together, the evidence from recent studies suggests that Ly$\alpha$ visibility during the EoR may evolve differently in UV bright and faint galaxies \citep[e.g.,][]{Oesch2015a, Zitrin2015a, Roberts-Borsani2016a, Stark2017a, Zheng2017a, Mason2018b}.  This is explained as UV bright galaxies are likely located in highly ionized bubbles which were created by a wealth of ionizing photons produced by those galaxies as well as a potentially larger number of nearby fainter galaxies \citep[][Larson et al. 2021 in preparation]{Finkelstein2019b}.  \cite{Endsley2021b} provide additional evidence for accelerated reionization around massive galaxies.  Particularly, they find a higher Ly$\alpha$ detection rate at $z\simeq7$ from massive galaxies with strong [\ion{O}{3}]+H$\beta$ emission, which reflects enhanced ionizing photo production rates \citep[e.g.,][]{Roberts-Borsani2016a, Tang2019a, Tang2021a, Endsley2021a}, arguing for higher Ly$\alpha$ equivalent-widths (EWs) from the strong [\ion{O}{3}]+H$\beta$ emitting population as similar as shown at lower redshifts of $z\simeq2$ -- $3$ \citep{Tang2021b}.  This may be related to the galaxies' specific star-formation rates (sSFRs) and ionization \citep[][Papovich et al., in prep]{Backhaus2021a}{, although this remains tenuous as there is not yet any conclusive evidence for a significant enhancement of Lyman continuum escape fraction found for galaxies with higher [\ion{O}{3}]+H$\beta$ galaxies at $z\sim3$ \citep{Saxena2021a}.}  Therefore, it is prudent to look for indications that evolution of Ly$\alpha$ emission in galaxies depends on UV luminosity in the epoch of reionization.

In this study, we present an analysis of the  CANDELS  Ly$\alpha$  Emission  At  Reionization (CLEAR) observations in the CANDELS GOODS fields \citep{Grogin2011a, Koekemoer2011a}.  As discussed below, CLEAR uses slitless IR spectroscopy from the HST WFC3.  Therefore, the CLEAR data provide (unbiased) constraints on Ly$\alpha$ emission in galaxies at $6.0<z<8.2$. This allows us to constrain the evolution of the Ly$\alpha$ EW in these galaxies. We use these data to study the evolution as a function of both redshift and galaxy UV absolute magnitude. Section 2 describes the CLEAR high-$z$ galaxy dataset (data reduction, sample selection, and emission-line and continuum-detection search) and the additional Keck DEIMOS and MOSFIRE observations from the Texas Spectroscopic Search for Ly$\alpha$ Emission at the End of Reionization Survey \citep{Jung2018a, Jung2020a}, finalizing the $6.0<z<8.2$ galaxy dataset for the Ly$\alpha$ analysis. In Section 3, we present our results, which includes the measurements of the Ly$\alpha$ EW distribution and the IGM transmission to Ly$\alpha$.  We summarize and discuss our findings in Section 4.  In this work, we assume the Planck cosmology \citep{Planck-Collaboration2016a} with $H_0$ = 67.8\,km\,s$^{-1}$\,Mpc$^{-1}$, $\Omega_{\text{M}}$ = 0.308, and $\Omega_{\Lambda}$ = 0.692.  The Hubble Space Telescope (HST) F435W, F606W, F775W, F814W, F850LP, F105W, F125W, F140W, and F160W bands are referred to as $B_{435}$, $V_{606}$, $i_{775}$, $I_{814}$, $z_{850}$, $Y_{105}$, $J_{125}$, $JH_{140}$ and $H_{160}$, respectively.  All magnitudes are given in the AB system \citep{Oke1983a}, and all errors presented in this paper represent 1$\sigma$ uncertainties (or central 68\% confidence ranges), unless stated otherwise. {All EWs discussed in this paper represent the rest-frame values, unless defined otherwise.}

\section{Data: $6.0<z<8.2$ Galaxies}
\subsection{\normalsize CLEAR \textit{HST}/grism Survey}
The CLEAR Experiment is a cycle 23 HST observing program (Program GO-14227, PI: C.~Papovich), which observed 12 fields  in the CANDELS GOODS fields to 10 to 12-orbit depth with the G102 grism in the HST/WFC3 camera. Each field was observed at 3 position angles (separated by $>$10 degrees) to properly correct galaxy spectra from contamination. The goal of the CLEAR survey is to measure the distribution of Ly$\alpha$ emission in galaxies during the epoch of reionization at $6.0 < z < 8.2$. {The feasibility of detecting Ly$\alpha$ emission from HST/WFC3 grism observations has been proved in previous studies, such as \cite{Schmidt2016a} from the GLASS survey \citep{Treu2015a}, and \cite{Tilvi2016a} and \cite{Larson2018a} from the FIGS survey \citep{Pirzkal2017a}.} The dataset also provides constraints on the stellar populations of $1 < z < 2$ galaxies.  Previous work on the metallicities, ages, and formation histories of massive galaxies at $1 < z < 2$ has been published in \cite{Estrada-Carpenter2019a,Estrada-Carpenter2020a}, and the gas-phase metallicity gradients of star-forming galaxies is investigated in \cite{Simons2020a}. Also, \cite{Cleri2020a} studied Paschen-$\beta$ as a star-formation rate indicator in low redshift galaxies, using the CLEAR dataset.

The dataset has been extended to include all publicly-available HST/WFC3 G102 and G141 grism observations in the CLEAR fields. For processing the grism observations, the grism redshift and line analysis software {\sc Grizli} \citep{Brammer2019a} has been utilized, which retrieves the raw observations and performs astrometric alignment, contamination modeling, extracting spectra, and fitting continuum and emission line models.  The full details of the grism data reduction and spectral extractions are described in \cite{Estrada-Carpenter2019a, Estrada-Carpenter2020a} as well as \cite{Simons2020a}.

While the primary CLEAR spectral extractions were made for galaxies based on the 3D-HST GOODS catalog \citep{Skelton2014a}, we extracted the CLEAR spectra of high-$z$ galaxies based on the updated HST CANDELS photometry and its segmentation maps from \cite{Finkelstein2021a}.  The photometric selection of high-$z$ candidate galaxies was done following the criteria described in Section 3.2 in \cite{Finkelstein2015a}.  In brief, the selection is based on the full photometric redshift probability distribution functions of $P(z)$ calculated by {\sc EAZY} \citep{Brammer2008a} rather than simply using the best-fit redshifts. This includes, for example, the integral of $P(z)$ under the primary redshift peak must be $>$70\% of the total $P(z)$ \cite[for the full detail, refer to][]{Finkelstein2015a}. Additionally, visual inspection was performed for removing any artifacts (e.g., diffraction spikes, nearby bright sources) and checking the quality of photometry.  This resulted in 180, $6.0<z_{\text{phot}}<8.2$ galaxies in the CLEAR fields.  However, checking the quality of the extracted spectra, in total 148 CLEAR spectra were collected after removing 32 spectra due to technical issues in treating the removal and corrections from contaminating spectrum from nearby galaxies using {\sc Grizli} (e.g., we removed objects \deleted{from the same} where residuals from contaminating spectra from nearby bright galaxies were too severe compared to the expected continuum or line flux for our objects).

\subsection{Emission Line and Continuum Detection in CLEAR}
In order to search Ly$\alpha$ emission lines or continuum objects detected from these high-$z$ targets, we utilized the grism spectroscopy from the full sample of the CLEAR spectra of 148, $6.0<z_{\text{phot}}<8.2$ galaxies. In this subsection, we describe the process we used to to determine if any emission lines or continuum spectra are present in the spectra of each galaxy. 

\subsubsection{Emission-Line Search}
We first found 24 galaxies from the {\sc Grizli} run with candidate emission lines detected at significance $>$3$\sigma$ from their co-added 1-dimensional (1D) spectra.  For these galaxies, we inspected their 2-dimensional (2D) spectra and those of nearby objects to see if the emission lines are a result of contamination.  Additionally, we attempted to secure detections from multiple PAs, requiring $>$2$\sigma$ detections in the data taken at different orientations. This assists in our effort to rule out artefacts in the data including contamination from nearby objects (as these will be different in each PA). 

Our emission-line search identified only one tentative Ly$\alpha$ detection at high confidence. This object (GN4\_5461276) shows an emission line that would correspond to {$z_{\text{grism}}=6.513\pm0.005$} with a detection significance of $\simeq 4\sigma$.  All other candidates showing potential Ly$\alpha$ emission either correspond to residuals of nearby contaminating sources (16 candidates) or possible spurious sources that show the emission in only a single PA (7 candidates).  In the case of the  emission line detected in GN4\_5461276,  the HST ACS $z_{850}$-band data challenges this measurement: if the emission line (interpreted as Ly$\alpha$) is real, then it should contribute $37\pm8$ nJy to the $z_{850}$-band flux.  However, the measured $z_{850}$-band flux of this galaxy in the HST/CANDELS observations is only $f_\nu(z_{850}) = 6.3\pm8.3$ nJy.  Thus, it may require further observational evidence for validating this tentative detection as Ly$\alpha$, and we do not include it in the remainder of our analysis for the Ly$\alpha$ EW distribution and the IGM transmission.  Instead, we discuss the details of this tentative detection in the Appendix.  Additionally, we suggest that it would be crucial to include an automated and improved removal process of contaminating emission-line sources in slitless spectroscopic surveys. This will be very important in the study of emission lines in data from the next generation space telescopes such as the James Webb Space Telescope (JWST), Euclid, or Nancy Grace Roman Space Telescope (NGRST).

\subsubsection{Continuum-Detection Search} 
We also inspected 42 sources with potential continuum detections identified from our visual inspection of the coadded 1D and 2D grism spectra. We vetted these further to determine if they are bona fide continuum sources.  First, we estimated $Y_{105}$, $J_{125}$, or $H_{160}$ magnitudes from the CLEAR spectra of the 42 candidates by applying the HST filter transmission curves to the reduced {co-added} 1D CLEAR spectra and compared them to the known HST magnitudes {of the sources} from the HST/CANDELS photometry from \cite{Finkelstein2015a}.  We required agreement within the 2$\sigma$ uncertainties in this comparison.  The continuum flux is estimated by summing all fluxes after the continuum break, and the continuum break is located where the estimated signal-to-noise (S/N) value of the continuum flux is maximized.  Thus, we required a secure (i.e., $>$3$\sigma$) continuum detection redward the Lyman-$\alpha$ continuum break measured from the {\sc Grizli}-extracted 1D spectrum.

From the selection above, we obtained 15 candidates which satisfy the continuum selection criteria.  On this subset, we then performed additional visual inspections of the galaxies' 2D and 1D spectra taken at the different PAs to check for possible contamination from nearby sources.  From these inspections, we identified contaminating sources in 6 of these 15 candidates, and no other candidates display clear continuum detection in multiple PAs due to the limited observing depth (as the depth in each individual PA is $\sim$1/3rd the exposure time of the total).  In general, therefore, the depth of the data is insufficient for reliable detection of the continuum in our target galaxies of $J_{125}\gtrsim26$ at individual PA-depth ($\sim$3 orbits).  However, we identify possible continuum in one source (GN2\_23734), which we discuss in the Appendix.  This galaxy shows possibly faint continua in both G102 and G141 spectra.  However, confirmation of this source would require deeper observations to locate the continuum break confidently.

While the current analysis of the slitless grism observations has achieved a great success in removing contamination from moderate brightness objects, this remains a factor when studying emission in faint galaxies. For example, even 1\% residuals from the spectra of contaminating sources leave a signal as bright as that from objects of interest for magnitude differences of $\Delta m$=5 mag (as will be the case for 27th--magnitude galaxies near galaxies brighter than 22nd magnitude).  Mitigating the effects of contamination will be important as detecting galaxy continuum breaks provides a means to confirm high-$z$ galaxies without Ly$\alpha$:  this will be particularly useful into the epoch of reionization where observable Ly$\alpha$ emission may be less frequent \citep[e.g.,][]{Rhoads2013a, Watson2015a, Oesch2016a}.

In summary, we do not identify any convincing Ly$\alpha$ emission or continuum detected galaxies from the CLEAR spectra. However, this in itself is an important finding as the CLEAR data for 148 $6.0<z_{\text{phot}}<8.2$ candidate galaxies still provides constraints on the evolution of the Ly$\alpha$ EW distribution.  This is because some bright Ly$\alpha$ emission lines could (even \textit{should}) have been detected with the CLEAR observing depth if the Ly$\alpha$ EW distribution does not evolve from $z \lesssim 6$ to $z \gtrsim 6$.  The non-detections in CLEAR rule out the existence of large EW Ly$\alpha$ sources in the CLEAR sample. We utilize the CLEAR spectra of these 148 $6.0<z_{\text{phot}}<8.2$ candidate galaxies for the remainder of this study to set constraints on the Ly$\alpha$ EW distribution.

\subsection{Supplemental data from Keck DEIMOS and MOSFIRE Observations}
The Texas Spectroscopic Search for Ly$\alpha$ Emission at the End of Reionization Survey \citep{Jung2018a, Jung2019a, Jung2020a} obtained deep spectroscopic data of over 120 $z>6$ galaxies in the GOODS-N fields (65 galaxies from DEIMOS and 62 galaxies from MOSFIRE observations). These data provided the largest {number} of confirmed Ly$\alpha$ emitters at $z>7$ with which to constrain the IGM \ion{H}{1} fraction at $z\sim7.6$ \citep{Jung2020a}. Using these data, \cite{Jung2020a} noted tentative evidence that the Ly$\alpha$ EW distribution at $z > 7$ depends on the galaxies' UV luminosities.  Here we use the CLEAR sample to explore this finding, as the CLEAR data provides constraints on Ly$\alpha$ emission for a larger sample of faint galaxies.

CLEAR aims to measure the Ly$\alpha$ EW distribution during the reionization era at $z>6$.  Also, CLEAR is a blind survey with slitless spectroscopy, providing galaxy spectra for a wide dynamic range of UV magnitudes.  Even non-detections in CLEAR  are still highly constraining, placing upper limits of Ly$\alpha$ visibility \citep{Treu2012a, Jung2018a, Mason2019a}.   The combination of the CLEAR data and Keck data then provided important, complimentary data.  This allows us to explore the evolution of the Ly$\alpha$ EW distribution and its dependence on UV absolute magnitude ($M_\text{UV}$) at higher significance than possible with either dataset separately.

\begin{figure*}[ht]
\centering
\includegraphics[width=1.05\columnwidth]{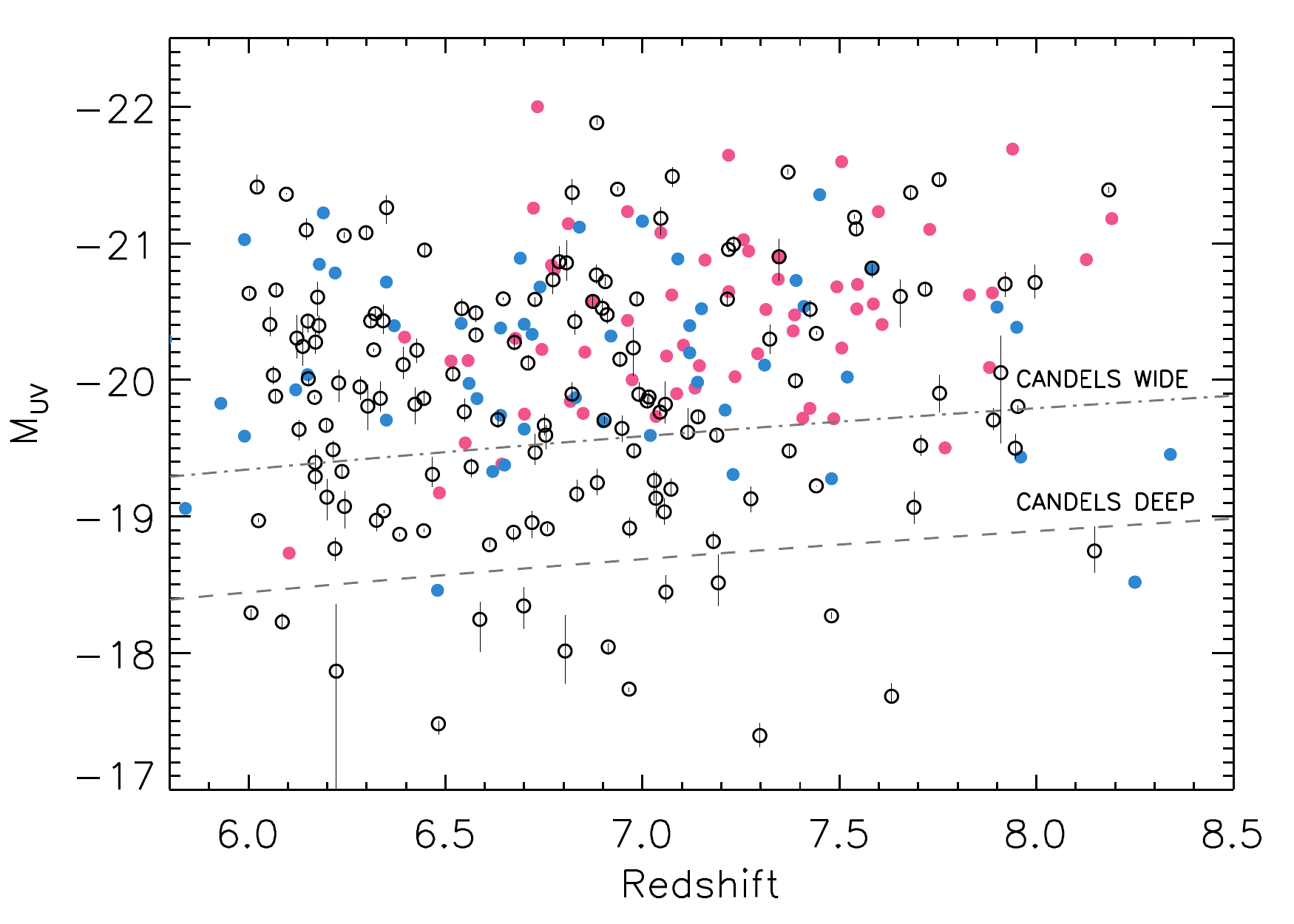}
\includegraphics[width=1.05\columnwidth]{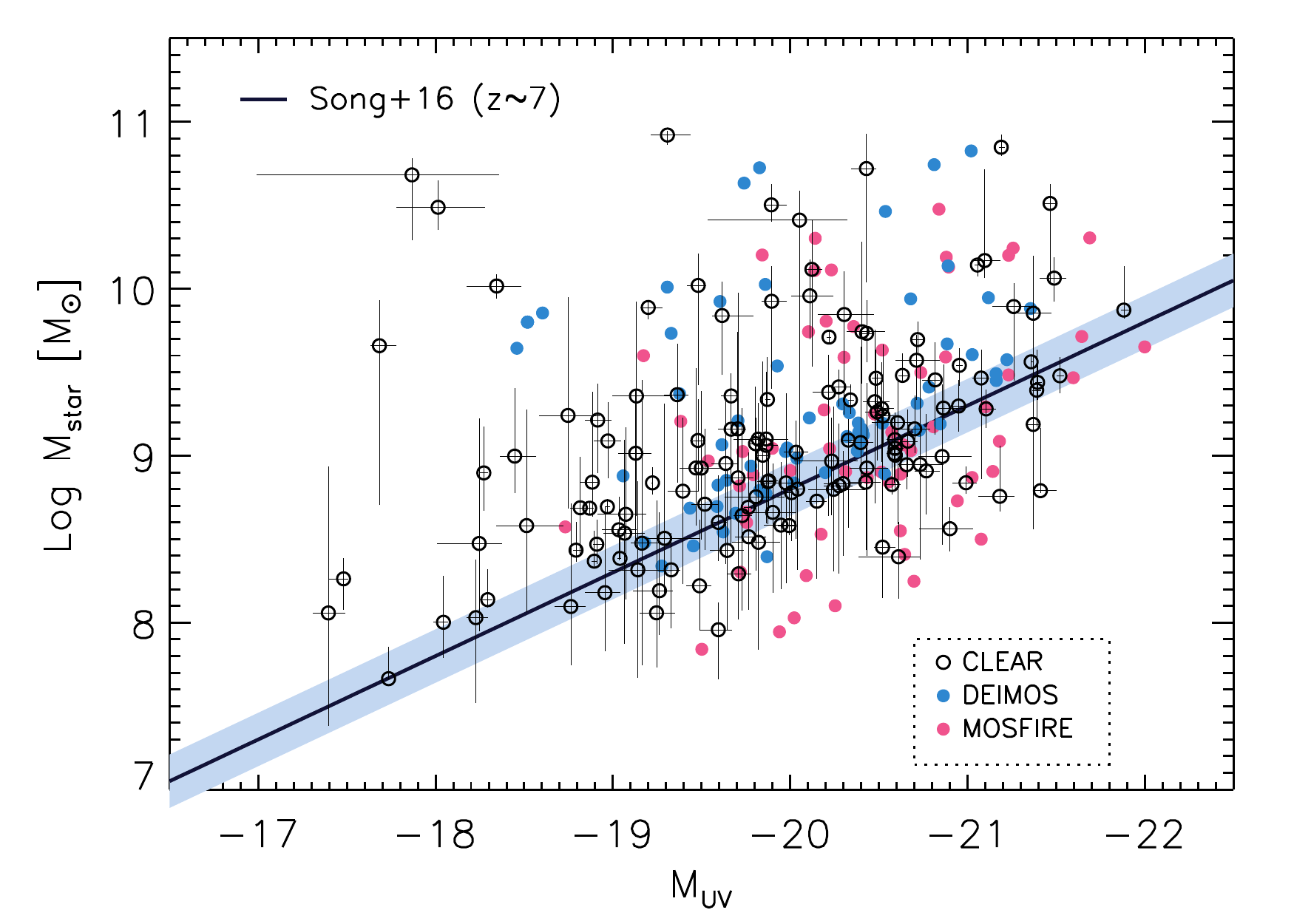}
\caption{The left panel shows the $M_{\text{UV}}$ distribution of target galaxies in the CLEAR dataset as a function of redshift, and the $M_{\text{star}}$--$M_{\text{UV}}$ relation in the right panel.  All 148 galaxies in the CLEAR sample are shown as open circles while the targets from the Keck observations are denoted as filled blue (DEIMOS: 65 galaxies) and red (MOSFIRE: 62 galaxies) circles.  A wide dynamic range of UV magnitudes in the CLEAR sample apparently show an unbiased nature of the blind survey, compared to the targeted Keck observations.  The blue solid line in the right panel shows the fiducial $z\sim7$ relation of \cite{Song2016b} with the shaded area of its uncertainty.}
\label{fig:m_z}
\end{figure*}

\subsection{Physical Properties of Combined Sample}
We derived the properties of our 148 sources in CLEAR by performing spectral energy distribution (SED) fitting with the \cite{Bruzual2003a} stellar population model, following \cite{Jung2019a, Jung2020a}.  We utilized the HST ACS and WFC3 broadband photometry ($B_{435}$, $V_{606}$, $i_{775}$, $I_{814}$, $z_{850}$, $Y_{105}$, $J_{125}$, $JH_{140}$ and $H_{160}$) in addition to Spitzer/IRAC 3.6$\mu$m and 4.5$\mu$m band fluxes.  {Here, we use photometric redshift measurements that have been updated in \cite{Finkelstein2021a} based on updated CANDELS photometry, including deeper $I_{814}$ and Spitzer/IRAC data (where those authors utilized a deblending technique to measure IRAC photometry more accurately using the HST images as priors). }
A \cite{Salpeter1955a} initial mass function was assumed with a stellar mass range of 0.1-100$M_{\odot}$, and we allowed a metallicity range of 0.005-1.0$Z_{\odot}$.  We explored a range of exponential models of star formation histories with exponentially varying timescales ($\tau=$ 10 Myr, 100 Myr, 1 Gyr, 10 Gyr, 100 Gyr, $-$300 Myr, $-$1 Gyr. $-$10 Gyr).  We use the \cite{Calzetti2001a} dust attenuation curve with $E(B-V)$ values ranging from 0 to 0.8 for modeling galaxy spectra and add nebular emission lines, adopting the \cite{Inoue2011a} emission-line ratio, as done in \cite{Salmon2015a}.  The IGM attenuation was applied to the SED models based on \cite{Madau1995a}.  

Physical properties were obtained from the best-fit models, minimizing $\chi^2$.  The uncertainties of physical parameters were estimated from SED fitting with 1000 Monte Carlo (MC) realizations of the perturbed photometric fluxes, according to their photometric errors for individual galaxies.  UV magnitudes ($M_{\text{UV}}$) of galaxies were calculated from the averaged fluxes over a 100\AA-bandpass (at 1450-1550\AA) from the best-fit models, which were not dust-corrected.  In our SED fitting, we fixed redshifts with the best-fit photometric redshifts calculated by {\sc EAZY}.

In the left panel of Figure \ref{fig:m_z} we present the $M_{\text{UV}}$ distribution of our CLEAR targets as a function of redshift with the existing Keck DEIMOS and MOSFIRE targets from \cite{Jung2018a, Jung2020a}.  The black open circles show the CLEAR sample, and the blue and red filled circles denote targets from DEIMOS and MOSFIRE observations, respectively.  The CLEAR galaxy sample does not show any noticeable bias, randomly distributed over a range of $-18 \gtrsim M_{\text{UV}} \gtrsim -22$, giving more UV faint ($M_{\text{UV}}>-19$) targets than the Keck observations.  However, the Keck observations are much deeper than CLEAR in general, providing significantly lower EW limits for those targeted galaxies.  Figure \ref{fig:m_z} also shows the stellar mass ($M_{\text{star}}$) -- $M_{\text{UV}}$ distribution of our CLEAR galaxies in the right panel, compared to the fiducial relation derived at $z\sim7$ by \cite{Song2016a}.  Similar to the Keck sample, most of the CLEAR targets are placed on the fiducial $M_{\text{star}}$--$M_{\text{UV}}$ relation at $z\sim7$ of \cite{Song2016a} with no significant selection bias from the typical high-$z$ galaxy population, although there are a few UV faint and massive galaxies located above the fiducial relation. 

\section{Ly$\alpha$ EW Distribution and IGM Transmission}
\subsection{Measuring the Ly$\alpha$ EW Distribution at $6.0<z<8.2$}

The Ly$\alpha$ EW distribution is commonly described by a declining exponential form, 
\begin{equation}{\label{eqn:Wdist}}
P(\text{EW})\propto\text{exp}^{-\text{EW}/W_0},
\end{equation}
where EW is the Ly$\alpha$ rest-frame EW, and $W_0$ is a $e$-folding scale width \citep[e.g.,][]{Cowie2010a}. \cite{Jung2018a} introduced an improved methodology of measuring the Ly$\alpha$ EW distribution, motivated by earlier studies \cite[e.g.,][]{Treu2012a, Treu2013a}, which utilizes galaxies without detected emission lines as well as Ly$\alpha$ detected objects. Following \citeauthor{Jung2018a}, we constructed a template of an expected number of Ly$\alpha$ emitters as a function of detection significance, by accounting for all types of data incompleteness including instrumental wavelength coverage, the wavelength-dependent Ly$\alpha$ detection limit, the UV-continuum flux, and the photometric redshift probability distribution function, $P(z)$, based on the combined dataset of the CLEAR spectra. We then applied this to the CLEAR dataset, and combined these results with those from our previous work using the Keck/DEIMOS+MOSFIRE observations in \cite{Jung2018a,Jung2019a,Jung2020a} to measure the Ly$\alpha$ EW distribution.

To calculate the expected number of Ly$\alpha$ emitters for a given model we did the following. We first estimated the detection sensitivity for an unresolved emission line in the grism spectrum for each object.  The detection limit was inferred through Monte Carlo simulations with the 1D spectra by assigning a mock Ly$\alpha$ emission line to the reduced 1D spectra for all objects.  We recovered the line flux of this mock emission line with its error from the Gaussian line fitting with the IDL MPFIT package \citep{Markwardt2009a}.  This mock emission line was created to have an intrinsic line profile equal to the best-fit line profile from one of the highest-S/N Ly$\alpha$ emission detected in z7\_GND\_42912 from the MOSFIRE observations of \cite{Jung2019a, Jung2020a}, and we smoothed the line profile to match the G102 resolution ($\sim$45\AA).  We note that the choice of the emission line profile has a negligible effect given the relatively low resolution of the G102 grism data ($R\sim 200$).   {We consider our targets to be point-like sources in our 1D mock emission simulation.  This assumption is appropriate as our target galaxies at this redshift are small, with a typical effective radius ($R_{\text{eff}}$) of  $R_{\text{eff}}<1$ kpc in general (or $\lesssim$3 pixels at the resolution of \textit{HST}/WFC3) and there is evidence that these kinds of galaxies are even smaller at higher redshifts \citep{Shibuya2019a}.  Additionally, in {\sc Grizli}, our 1D spectra are extracted using the source segmentation maps from the photometric catalog, thus the extracted 1D spectra include the contribution estimated from the (rest-frame UV) morphology as measured by the $Y_{105}$ band.  We may not capture the contribution of Ly$\alpha$ from extended halos in 1D spectra \citep[e.g,][]{Leclercq2017a, Wisotzki2018a}.  However, an appropriate modeling of the extended component of Ly$\alpha$ at this redshift is beyond the scope of this study.  In future, for example, JWST NIRSpec IFU observations will resolve the extended component of Ly$\alpha$.}

\begin{figure}[!t]
\centering
\includegraphics[width=1.04\columnwidth]{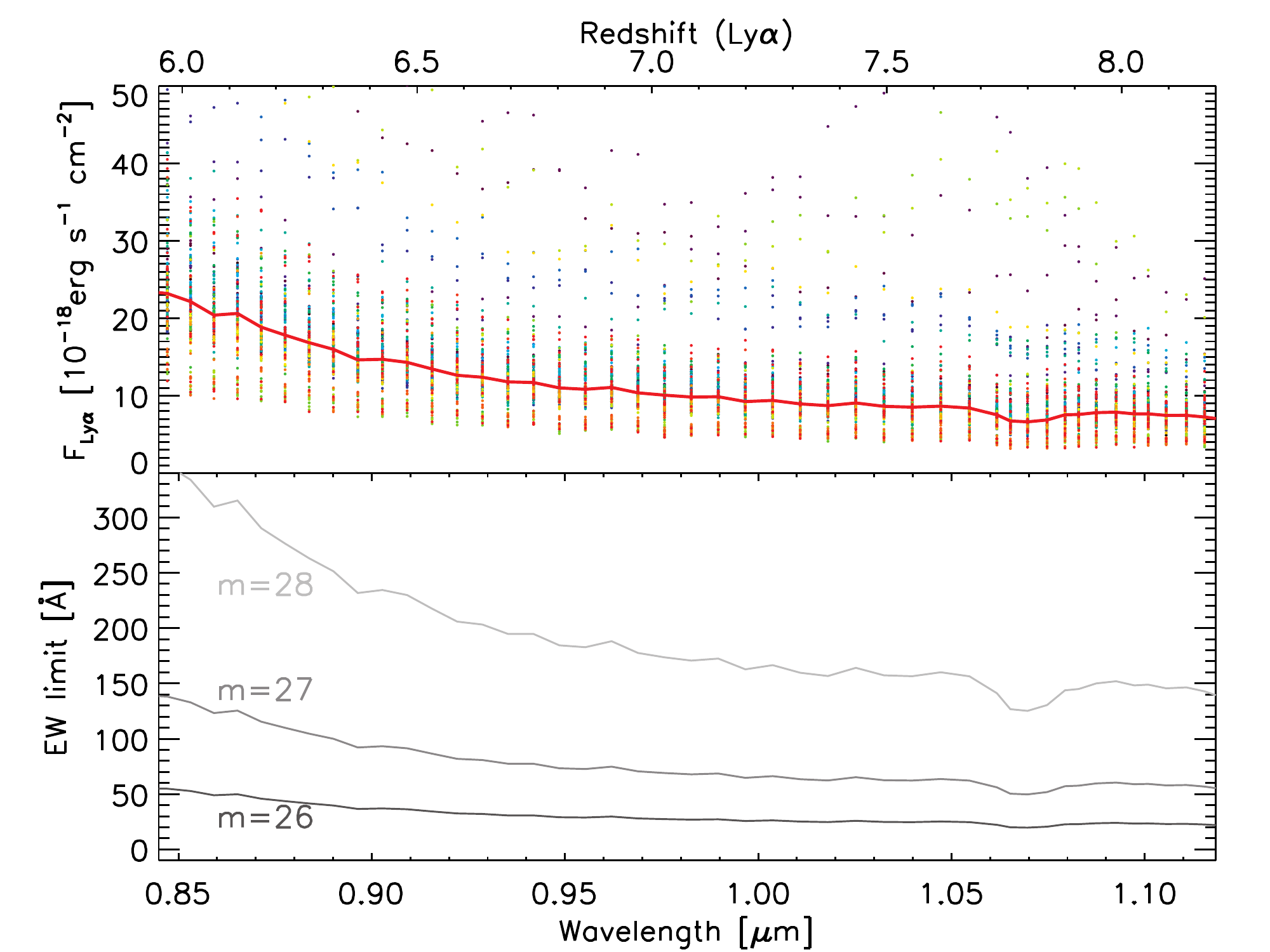}
\caption{(Top) $3\sigma$ emission-line detection limits across the wavelengths corresponding to the WFC3 G102 redshift coverage of Ly$\alpha$ at $6.0<z<8.2$. The red curve represents the median detection limit, and the detection limits of the 148 individual galaxies analyzed here are shown colored dots.  The $3\sigma$ detection limit reaches down to below $\sim$1$\times10^{-17}$ erg\ s$^{-1}$\ cm$^{2}$ around 1.0 - 1.1$\mu m$ where the G102 grism transmission is high. (Bottom) The median 3$\sigma$ rest-frame EW limits as a function of wavelength for varying apparent UV magnitudes (as labeled).}
\label{fig:lya_emission_limit}
\end{figure}

From these simulation emission lines we can derive emission-line flux limits for the CLEAR G102 data.  The top panel of Figure \ref{fig:lya_emission_limit} shows the 3$\sigma$ detection limits over the wavelength range of the G102 grism, which corresponds to a redshift range of for Ly$\alpha$ emission at $6.0 \lesssim z \lesssim 8.2$.  In the figure, the colored dots show the detection limits for all the 148 individual galaxies in our sample; the red curve shows the median detection limit for all spectra.  The bottom panel of Figure \ref{fig:lya_emission_limit} shows the corresponding (median) 3$\sigma$ rest-frame EW \deleted{(REW)} limits for galaxies as a function of apparent UV magnitude. The EW limit{s were} estimated by dividing the median 3$\sigma$ emission-line flux limits ($f_{3\sigma\text{--limit}}$) by the continuum level ($f_{\text{cont}}$) and ($1+z_{\text{Ly}\alpha}$) at each corresponding wavelength as below.

\begin{equation}
    \text{EW}_{3\sigma \text{--limit}} \text{ [\AA]} =\frac{f_{3\sigma\text{--limit}}}{f_{\text{cont}}(1+z_{\text{Ly}\alpha})}
\end{equation}

After detection limits estimated, we constructed a template of an expected number of detected Ly$\alpha$ emission lines as a function of detection significance.  The template construction was done in a Monte Carlo (MC) fashion, by simulating mock Ly$\alpha$ emission lines for entire target galaxies, following \cite{Jung2018a, Jung2020a}.  The MC simulation of individual mock emission lines has three main steps: (i) we allocate the wavelength of the simulated Ly$\alpha$ emission, randomly taken based on an object's $P(z)$; (ii) at each redshift, we simulate a Ly$\alpha$ line flux given the galaxy's magnitude and EW, where the latter is drawn from the distribution in Equation~\ref{eqn:Wdist} over a range for the assumed Ly$\alpha$ EW scale width, $W_0=$ 5-200~\AA;  (iii) finally, we measure a detection significance for this simulated emission line based on the precomputed emission-line detection limits taken from above.  For each choice of $W_0$, we carried out 1000 sets of MC simulations of the mock emission lines for the entire target galaxies, which produces a distribution of the expected number of Ly$\alpha$ detections as a function of detection significance.

\begin{figure*}[!t]
\centering
\includegraphics[width=0.8\textwidth]{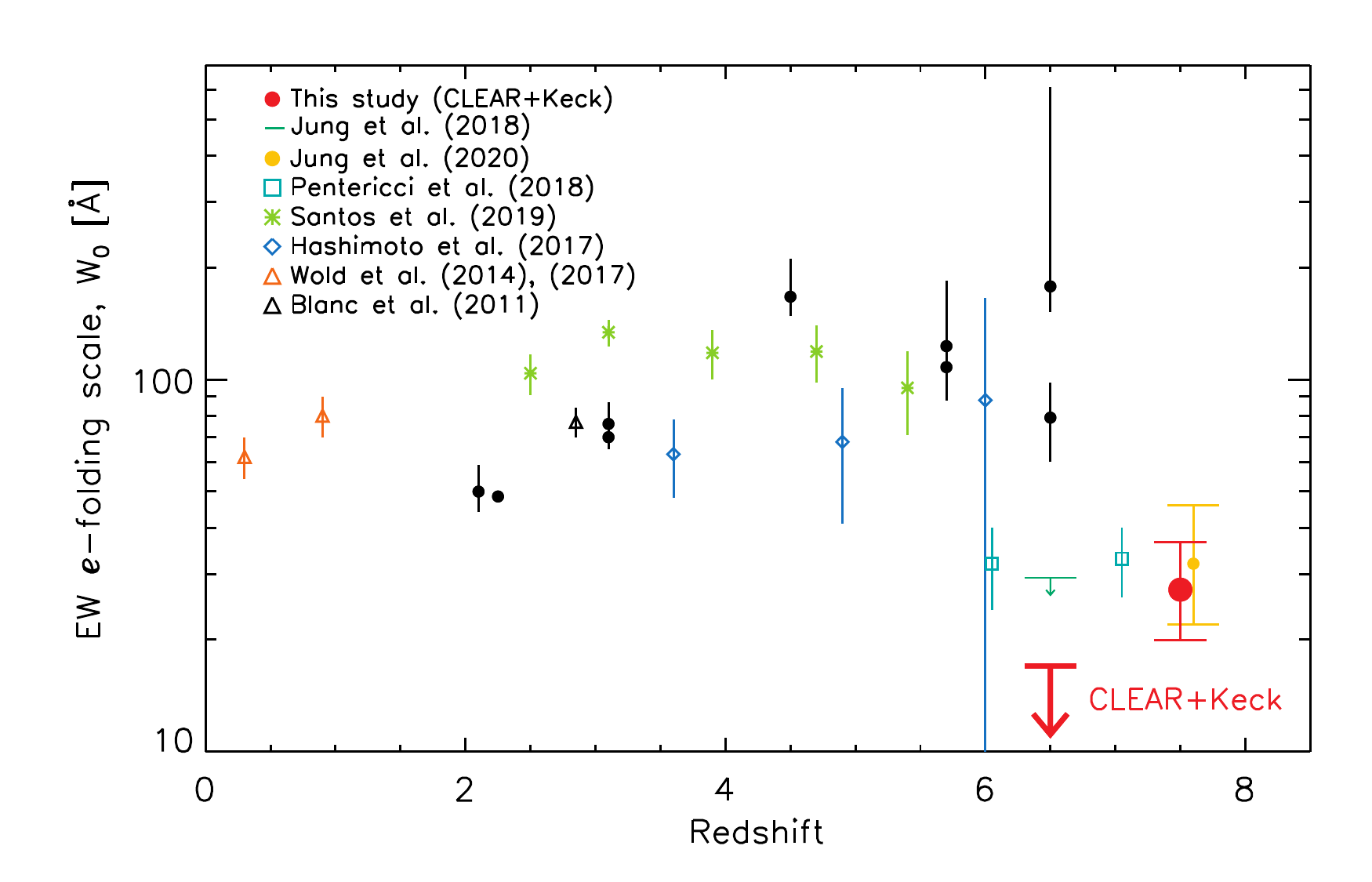}
\caption{Similar to Figure 11 in \cite{Jung2020a}: the redshift dependence of the Ly$\alpha$ EW $e$-folding scale width ($W_{0}$) up to $z\sim8$.  The figure includes a compilation of the previous $W_0$ measurements \citep{Gronwall2007a, Ouchi2008a, Nilsson2009a, Guaita2010a, Hu2010a, Kashikawa2011a, Ciardullo2012a, Zheng2014a, Wold2014a, Wold2017a, Hashimoto2017a, Jung2018a, Pentericci2018b, Santos2020a, Jung2020a}.  Our updated $W_0$ measurements with CLEAR are denoted by red, large symbols: horizontal bars at $z\sim6.5$ and a filled circle at $z\sim7.6$.  The $z\sim7.6$ measurement is displayed at $z=7.5$ to avoid an overlap with the \cite{Jung2020a} data point.  Our study confirms a decreased $W_0$ at $z>6$, whereas there is little/no redshift evolution of $W_0$ reported in the literature at $z<6$.  \deleted{Furthermore, we see little (or roughly constant) evolution in $W_0$ from $6.0 < z < 7.0$ to $7.0 < z < 8.2$.}}
\label{fig:ew_evolution}
\end{figure*}

Lastly, we fit the data for each galaxy to the suite of simulated templates over the range of redshift and emission line EW.  We then combined these results to construct a probability distribution function (PDF) for the $e$-folding scale width ($W_0$) of the Ly$\alpha$ EW distribution.  We utilized a Markov Chain Monte Carlo (MCMC) algorithm for this fitting. As counting the number of Ly$\alpha$ emission line detections is a general Poisson problem, we used a Poisson likelihood. We used the \citet{Cash1979a} statistic to describe the Poisson likelihood.  We then used a  Metropolis-Hastings MCMC sampler \citep{Metropolis1953a, Hastings1970a} to construct the PDF of $W_0$, generating 10$^5$ MCMC chains.  For more information on this methodology, see \cite{Jung2019a, Jung2020a}.

\subsection{The Evolution of Ly$\alpha$ EW at $z>6$}

\cite{Jung2018a} constrained the characteristic $e$-folding scale width $W_0$ of the Ly$\alpha$ EW distribution at $6<z<7$ using data from Keck/DEIMOS for a sample of galaxies. They measured 1$\sigma$ and 2$\sigma$ upper limits of $W_0 <$ 36\AA\ and $<$125\AA, respectively.   Similarly, \cite{Jung2020a} measured $W_0=32^{+14}_{-9}$\AA\ for galaxies at $7.0 < z < 8.2$ using data from Keck/MOSFIRE observations. These values are shown in Figure \ref{fig:ew_evolution}.

Here, we report new measurements on the evolution of the Ly$\alpha$ EW from the combined analysis of the CLEAR and Keck datasets. The CLEAR data provide important constraints on the evolution of Ly$\alpha$ as the lack of detections implies strong evolution in the number of sources with high Ly$\alpha$ EWs at these redshifts. The updated constraints improve the constraint on the EW $e$-folding scale width, $W_0$, to be $<$17~\AA\ (1$\sigma$) and  $<$28~\AA\ (2$\sigma$) at $6.0<z<7.0$ and $W_0 = 27^{+10}_{-7}$~\AA\ at $7.0<z<8.2$ (these are shown as red, large symbols in Figure \ref{fig:ew_evolution}).  These measurements confirm the decline of $W_0$ already seen at $6.0<z<7.0$ compared to $z < 6$.  \deleted{We also report tentative evidence for little (or roughly constant) evolution in $W_0$ from $6.0 < z < 7.0$ to $7.0 < z < 8.2$.}

\begin{figure*}[t]
\centering
\includegraphics[width=1.05\columnwidth]{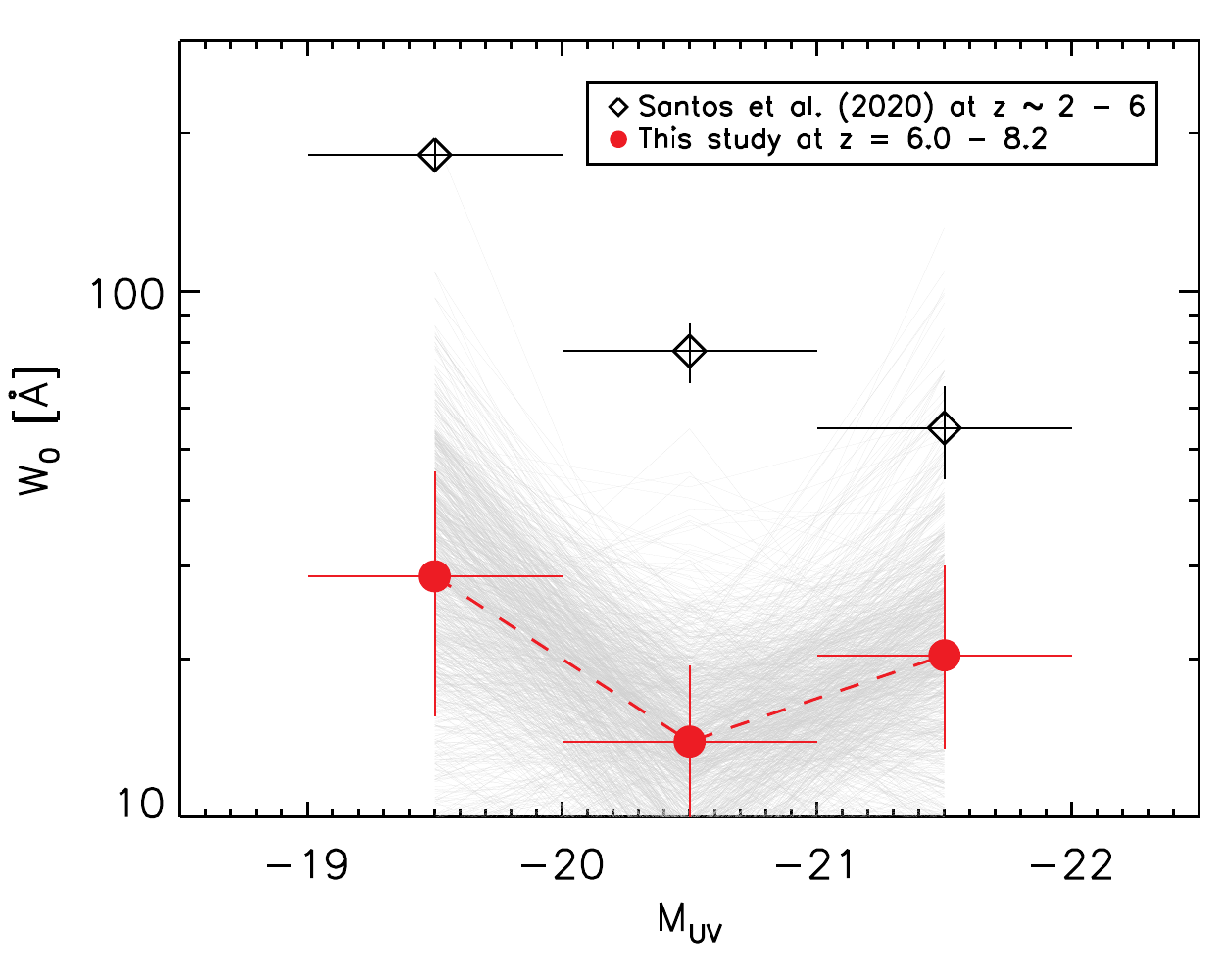}
\includegraphics[width=1.05\columnwidth]{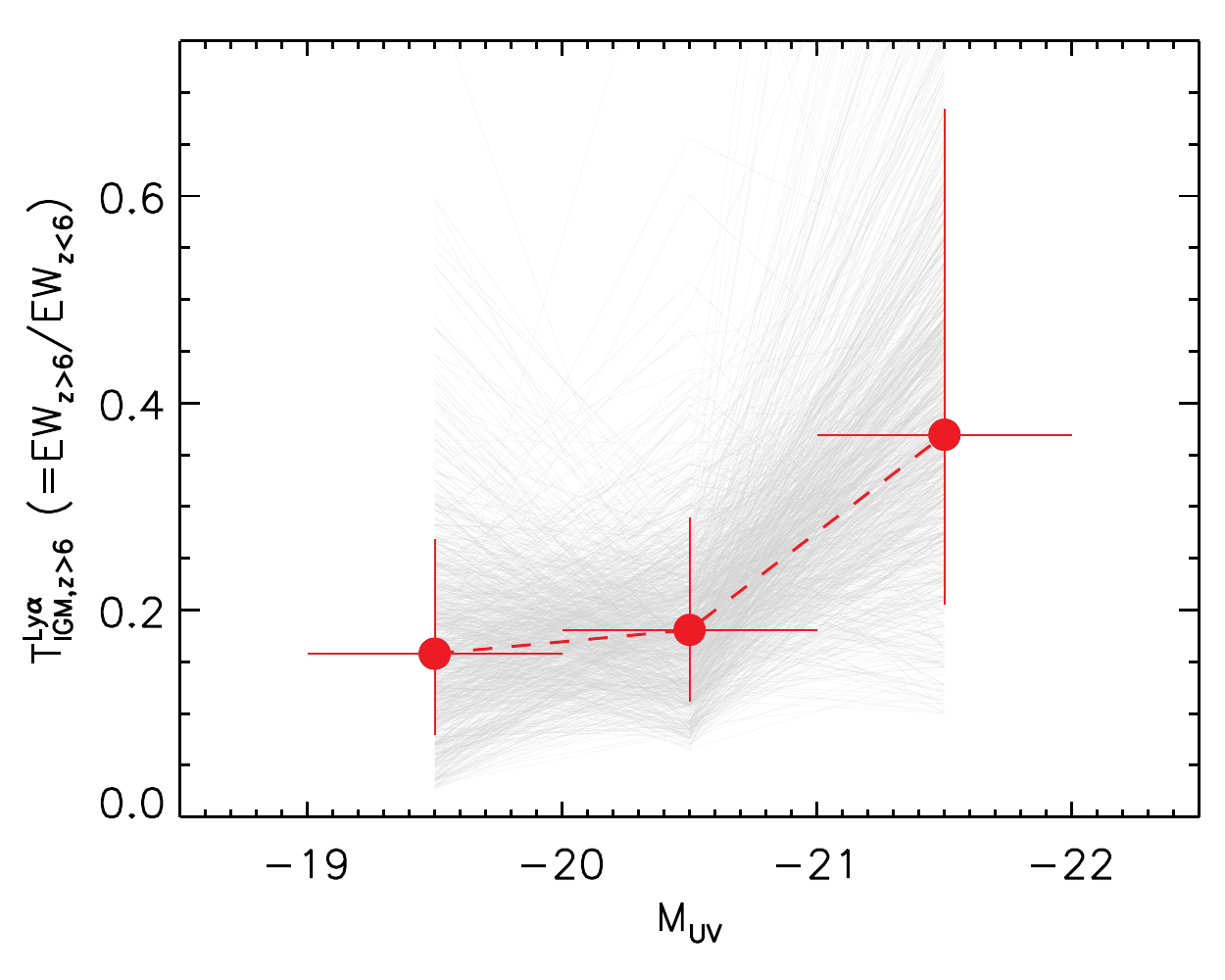}
\caption{(Left) Measurements of the $e$-folding scale width ($W_0$) of the Ly$\alpha$ EW distribution at different $M_{\text{UV}}$. Our $W_0$ measurements from this study (combining the CLEAR and Keck samples) at three $M_{\text{UV}}$ bins are shown as red filled circles, compared with those from lower-$z$ observations in \cite{Santos2020a}. Our $W_0$ measurement from the brightest UV magnitude galaxies show weaker evolution from $z<6$ to $z>6$ while those from fainter UV magnitude galaxies present a significant drop from $z<6$ to $z>6$. (Right) Ly$\alpha$ Transmission in the IGM ($T^{\text{Ly$\alpha$}}_{\text{IGM},z>6}$) at different $M_{\text{UV}}$. Our result indicates that the IGM transmission to Ly$\alpha$ during the reionization epoch is high from UV bright galaxies, compared to those from UV faint ones. {The gray lines in the backgrounds in both panels represent 1000 random draws from the $10^{5}$ MCMC chains that were generated for constructing the PDF of $W_0$.  We find that 76\% of the MCMC chains show an upturn in $W_0$ (or a greater $W_0$) in the UV brightest bin as shown in the left panel.  In addition, we find that 91\% of the chains suggest an increasing trend between the Ly$\alpha$ IGM transmission and the (increasing) UV luminosity in the right panel.}}
\label{fig:ew_muv}
\end{figure*}

\begin{deluxetable*}{@{\extracolsep{16pt}}ccccc}[!ht]
\setlength{\tabcolsep}{8pt}
\tablecaption{Ly$\alpha$ EW $e$-folding Scale Width ($W_0$) and the Ly$\alpha$ IGM Transmission ($T^{\text{Ly$\alpha$}}_{\text{IGM}}$) at Different $M_{\text{UV}}$\label{tab:EW_comparison}
} 
\tablehead{
\colhead{} & \colhead{}& \multicolumn{2}{c}{$W_0$} & \colhead{$T^{\text{Ly$\alpha$}}_{\text{IGM}}$} \\
\cline{3-4} \cline{5-5}
\colhead{$M_{\text{UV}}$ ($L_{\text{UV}}/L^{*}$\tablenotemark{a})} & \colhead{$N_{\text{galaxy}}$} & \colhead{$z\sim2-6$}  & \colhead{$6.0<z<8.2$} & \colhead{$6.0<z<8.2$} \\
\colhead{} & \colhead{} & \colhead{\cite{Santos2020a}}  & \colhead{This study} & \colhead{}
}
\startdata
{$-20<M_{\text{UV}}<-19$ ($0.15<L_{\text{UV}}/L^{*}<0.4$)\tablenotemark{b}} & 77 & {$178^{+13}_{-13}$\AA}  & {$29^{+16}_{-13}$\AA} & {$0.16^{+0.11}_{-0.08}$} \\  
{$-21<M_{\text{UV}}<-20$ ($0.4<L_{\text{UV}}/L^{*}<1.0$)} &  92 & {$73^{+10}_{-10}$\AA}  & {$14^{+5}_{-4}$\AA} &  {$0.18^{+0.11}_{-0.07}$} \\  
{$-22<M_{\text{UV}}<-21$ ($1.0<L_{\text{UV}}/L^{*}<2.5$)} &  30 & {$54^{+11}_{-11}$\AA}  & {$20^{+10}_{-6}$\AA} &  {$0.37^{+0.31}_{-0.17}$} \\
\enddata
\tablenotetext{}{
{\footnotesize
$^{a}$ $L_{\text{UV}}/L^{*} = 10^{0.4(M^{*}_{\text{UV}}-M_{\text{UV}})}$, where $M^{*}_{\text{UV}}$ is the characteristic magnitude of a typical galaxy with an approximate value of $M^{*}_{\text{UV}}\simeq-21$ at $z\sim7$ \citep{Finkelstein2015a}. \\
$^{b}$ Our sample in this faintest UV magnitude group could be incomplete in the CANDELS GOODS Wide fields due to the $\sim$1 magnitude shallower detection limits in the Wide fields than those in the CANDELS GOODS Deep fields as shown in Figure \ref{fig:m_z} (left). Thus, we also calculated $W_0$ in $-20<M_{\text{UV}}<-19$ excluding objects in the Wide fields, which provides $W_0=30^{+20}_{-15}$\AA. Although it is slightly increased from our fiducial value listed in the table, the change is insignificant and does not alter our conclusion.
}}
\end{deluxetable*}

\subsection{The Dependence of the Ly$\alpha$ EW on UV Magnitude} 

Observations at lower redshifts ($2 \lesssim z \lesssim 6$) show the the Ly$\alpha$ EW $e$-folding scale width ($W_0$) decreases with increasing UV luminosity \citep[e.g.,][]{Ando2006a, Stark2010a, Schaerer2011a, Cassata2015a, Furusawa2016a, Wold2017a, Hashimoto2017a, Oyarzun2017a, Santos2020a}.  In contrast, \cite{Jung2020a} reported tentative evidence that $W_0$ increases with increasing UV luminosity at $z > 6$ from the brightest galaxies ($M_{\text{UV}}<-21$).  Although the initial evidence was weak due to the large measurement errors, this finding is consistent with a picture of reionization where the IGM around brighter galaxies becomes ionized earlier than that of fainter objects, allowing for more efficient Ly$\alpha$-photon escape \citep[e.g.,][]{Zheng2017a, Mason2018a, Endsley2021b}.

CLEAR provides a means to improve the constraint on the evolution of $W_0$ as a function of absolute UV magnitude as the CLEAR data cover many more lower-luminosity objects.  We therefore recalculated $W_0$ from the combined dataset of CLEAR and Keck/DEIMOS+MOSFIRE observations in different bins of UV absolute magnitude ($M_\mathrm{UV}$).  
These $W_0$ measurements are listed in the third column in Table \ref{tab:EW_comparison} and shown in the left panel of Figure \ref{fig:ew_muv} as filled red dots.   
From this analysis, the CLEAR data improve the evidence that the $e$-folding scale width ($W_0$) of the Ly$\alpha$ EW distribution at $6.0<z<8.2$ shows a possible upturn of $W_0$ in the brightest UV magnitude bin, $-22<M_{\text{UV}}<-21$, although it is consistent with no upturn at 1$\sigma$ level. 
{The PDF of $W_0$ is constructed from $10^{5}$ MCMC chains as described in Section 3.1.  We then inspect what fraction of the MCMC chains have an upturn of $W_0$ in the brightest UV magnitude bin.  We find that 76\% of the MCMC chains show such an upturn (or a greater $W_0$).  In addition, we find that 91\% of the chains show an increasing trend between the Ly$\alpha$ IGM transmission and the (increasing) UV luminosity as shown in Figure \ref{fig:ew_muv}.  We discuss this further in the next subsection.}
This is in strong contrast with the measurements at $2 < z < 6$ that show $W_0$ decreases with increasing UV luminosity \citep[][shown as black open circles]{Santos2020a}, indicating that at $z > 6$ the Ly$\alpha$ emission for UV-brighter galaxies is less affected by the IGM compared to UV-fainter galaxies.

{We note that we cannot completely rule out the possibility of the redshift evolution of galaxy physical properties (e.g., ISM condition), which alters the Ly$\alpha$ properties significantly even without the IGM effect \citep{Hassan2021a}.  However, such decreasing trend of $W_0$ has been reported from most lower redshifts studies at $z<6$ (see references earlier in this section), and evolutionary features of Ly$\alpha$-emitting galaxies are not expected to be extreme particularly between $z<6$ and $z>6$.}

\subsection{Ly$\alpha$ Transmission in the IGM}

Our $W_0$ measurements indicate different evolution of the Ly$\alpha$ EW distribution between bright and faint galaxies into the epoch of reionization, $z \gtrsim 6$. As shown in Figure \ref{fig:ew_muv}, the $W_0$ values obtained for the the UV--fainter objects ($M_{\text{UV}}>-21$) at $ z> 6$ are substantially lower than those at lower redshift ($2 < z < 6$), whereas there is milder evolution in $W_0$ for the UV-brighter objects ($M_{\text{UV}}<-21$) from $z < 6$ to $z > 6$. 

The evolution of $W_0$ leads to constraints on the relative Ly$\alpha$ transmission, $T^{\text{Ly$\alpha$}}_{\text{IGM}}$
which we define as the ratio of observed--to--expected (intrinsic) rest-frame EW, EW$_{\text{obs}}$/EW$_{\text{int}}$, therefore,
\begin{equation}
T^{\text{Ly$\alpha$}}_{\text{IGM},z>6} = \frac{\text{EW}_{z>6}}{\text{EW}_{z<6}}
\end{equation}
where we take EW$_{z<6}$ as the intrinsic EW after the IGM became completely ionized.  Thus, we adopted the measurements of the EW distribution at $2 < z < 6$ from \cite{Santos2020a} as the intrinsic values and calculated the IGM transmission of Ly$\alpha$ at $z>6$ with our $W_0$ measurements.  

The estimated $T^{\text{Ly$\alpha$}}_{\text{IGM}}$ values are listed in the last column in Table \ref{tab:EW_comparison}. Tentatively, the fact that the IGM transmission for Ly$\alpha$ photons appears to be higher for the UV-brightest galaxies can be interpreted as evidence that the the IGM is highly ionized and more transparent around the brightest galaxies at $z > 6$, albeit this result will require further confirmation with larger and more sensitive datasets from future telescopes, such as JWST and NGRST.
In contrast, the decrease in IGM transmission for fainter galaxies ($M_{\text{UV}}>-21$) can be interpreted as evidence that the IGM is more neutral where Ly$\alpha$ photons suffer significantly more attenuation.  
This is consistent with a picture of the spatially inhomogeneous process of reionization where large ionized bubbles were often created by bright galaxies \citep[e.g.,][]{Zheng2017a, Castellano2018a, Jung2019a, Jung2020a, Hu2021a, Endsley2021a, Park2021a, Qin2021a}, and we expand on this interpretation in the next section. 

{We caution that our definition of the Ly$\alpha$ IGM transmission with the ratio of observed--to--expected (intrinsic) EW could be overly simplistic as it cannot describe variations of the Ly$\alpha$ IGM transmission depending the reionization history and ionized bubble sizes as well as Ly$\alpha$ emission properties of individual galaxies.  In principle, the Ly$\alpha$ transmission can be modeled using resionization simulations (and this has been the focus of several previous studies, e.g., \citealt{Mesinger2015a, Mason2018a, Weinberger2018a}).  
However, such approaches still depend on other assumptions (e.g., the reionization history, the evolution of ionized bubble sizes, and $M_{\text{UV}}$ -- halo mass relation).  Here, our intent is to provide results that are model-independent from the latest sets of Ly$\alpha$ observations combined with our empirical approach.  In future work, it will be very useful to extend these investigations to study how the Ly$\alpha$ IGM transmission varies depending on the reionization history and the characteristic ionized bubble sizes around galaxies \citep[e.g.,][]{Garel2021a, Park2021a, Smith2021a} as well as the intrinsic Ly$\alpha$ emission properties of galaxies during the epoch of reionization such as Ly$\alpha$ velocity offset \citep[e.g.,][]{Stark2017a, Hashimoto2018a, Hutchison2019a}.}

\begin{figure*}[!t]
\centering
\fbox{\includegraphics[width=0.82\paperwidth]{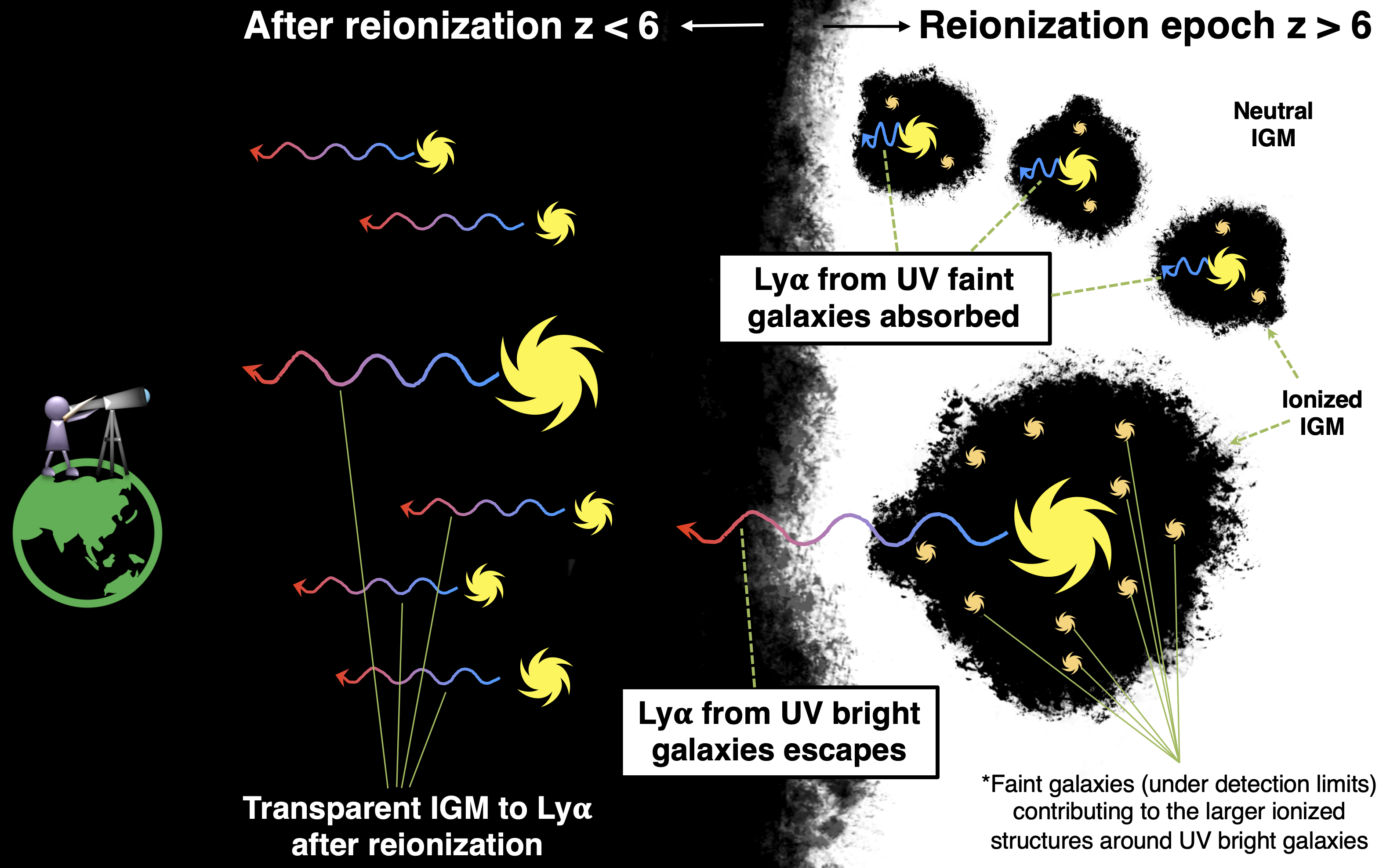}}
\caption{Cartoon summarizing the inhomogeneous process of reionization at $z>6$ universe consistent with the data and interpretation here.  Galaxies with higher UV luminosity have larger ionized bubbles around them.  This allows a higher Ly$\alpha$ transmission to the IGM. These UV bright galaxies are likely to be centered in overdense regions where they cluster with fainter galaxies (and these may be beyond current detection limits).  The combination of ionizing photons from the galaxies in the overdense regions may provide significant ionizing photon budget around UV bright galaxies \citep[][Larson et al. 2021, in prepration]{Finkelstein2019b, Endsley2021b}. In contrast, UV-fainter galaxies have more neutral gas around them (i.e., they sit in smaller ionized bubbles), thus the Ly$\alpha$ transmission to the IGM is suppressed.  At lower redshift, where the IGM is completely ionized the Ly$\alpha$ emission has overall high transmission for all galaxies.   In the figure, the darker (black) color shows a more ionized medium while a neutral gas is represented as bright (white) color. UV-brighter galaxies are shown with larger symbols.}
\label{fig:cartoon}
\end{figure*}

\section{Conclusions and Discussion}

We extracted HST slitless grism spectra of 148 high-$z$ galaxies in the GOODS fields from the CLEAR survey. Combining our high-$z$ dataset from CLEAR to the existing Keck Ly$\alpha$ observations from the Texas Spectroscopic Search for Ly$\alpha$ Emission at the End of Reionization Survey \citep{Jung2020a}, we provide an improved measurement of the $e$-folding scale width ($W_0$) of the Ly$\alpha$ EW distribution at $z>6$ and examine the dependence of $W_0$ on the UV luminosity ($M_{\text{UV}}$). Our findings are summarized as follows. 

\begin{enumerate}
\item We inspected the reduced CLEAR spectra of high-$z$ candidate galaxies in order to search for Ly$\alpha$ emission-lines or continuum-break detections.  The search finds only a tentative ($\sim 4\sigma$) Ly$\alpha$ detection from a {$z_{\text{grism}}=6.513\pm0.005$} galaxy and a continuum detection from a {$z_{\text{grism}}=6.2^{+2.5}_{-0.8}$} galaxy.  We discuss these more in the Appendix, although we cannot confirm their nature as high-$z$ sources from the current data due to the incomplete contamination removal and the shallow observing depth.  

\item With the combined dataset of CLEAR and the Keck observations, we compute the 1$\sigma$ (2$\sigma$) upper limit of $W_0$ at $<$17\AA\ ($<$28\AA) at $6.0<z<7.0$ and $27^{+10}_{-7}$\AA\ at $7.0<z<8.2$.  This is illustrated in Figure~\ref{fig:ew_evolution}.  This confirms a significant drop of $W_0$ previously suggested at $6.0<z<7.0$. \deleted{Furthermore, we see little or constant evolution in $W_0$ from $6 < z < 7$ to higher redshift $7 < z < 8.2$.}  The CLEAR data improved these constraints on the evolution of $W_0$ compared to the previous analysis using only the Keck data.  This is primarily a consequence of the spectral measurements for an increased number of galaxies, pushing to lower UV luminosities {($L_{\text{UV}}\approx0.025L^{*}$)} in the CLEAR sample. 

\item We constrained the $W_0$ values and estimated the IGM transmission of Ly$\alpha$ at $z>6$ as a function of UV absolute magnitude. This is shown in Figure~\ref{fig:ew_muv}. The Ly$\alpha$ EW distribution at $6.0<z<8.2$ show a possible upturn of $W_0$ in the UV brightest galaxies ($-22<M_{\text{UV}}<-21$) {at a 76\% confidence level}, in contrast with the strong decline in $W_0$ with increasing UV brightness for galaxies at $z<6$.  Consequently, the IGM transmission to Ly$\alpha$ at $z>6$ appears to be higher for the UV brightest galaxies {with 91\% confidence}. 
%

\item We interpret the evolution in $W_0$ as a function of UV luminosity as a change in the transmission of the IGM to Ly$\alpha$ at $z \gtrsim 6$ that depends on UV luminosity.  This is consistent with a picture where the IGM around the brightest galaxies 
 ($-22<M_{\text{UV}}<-21$) is largely ionized, allowing high Ly$\alpha$ transmission. 
 In this picture, the IGM is spatially inhomogeneous where large ionized bubbles are created by UV bright galaxies, while the IGM around fainter galaxies is more neutral, leading to strong attenuation of Ly$\alpha$ photons and lower transmission. 
 
\end{enumerate}

This picture builds on other recent observations that support an inhomogeneous nature of the IGM during reionization. We illustrate this picture in Figure~\ref{fig:cartoon}.   Studies of the IGM neutral fraction from Ly$\alpha$ emission show disparity in their measurements at $z\sim7$--8 \citep{Mason2018a, Mason2019a, Hoag2019a, Jung2020a}. Particularly, a higher Ly$\alpha$ detection rate reported in \cite{Jung2020a} suggests an overdense and highly ionized region in the GOODS-N field.  Recent studies of the Lyman-alpha damping wing in QSO spectra also provide a large scatter of the IGM neutral fraction at the same redshift \citep{McGreer2015a, Davies2018a, Wang2020a, Yang2020a}. One explanation for such disagreement between the IGM neutral fraction measurements is that reionization did not occur spatially-uniformly, but in an inhomogeneous way \citep[e.g.,][]{Finlator2009a, Pentericci2014a, Katz2019a}. 

In the same context, recent findings of clustered LAEs from bright galaxies in the middle phase of reionization \citep{Zheng2017a, Castellano2018a, Tilvi2020a, Jung2020a, Hu2021a, Endsley2021a} suggest high Ly$\alpha$ visibility from these bright galaxies, implying inhomogeneous reionization caused by individual/groups of bright galaxies. This picture is further supported by Ly$\alpha$ detections from two of the brightest $z >$ 8.5 galaxies \citep[][Larson et al. 2021 in preparation]{Zitrin2015a, Naidu2020a}, which implies that reionization proceeds more rapidly and/or earlier around brighter galaxies.

Our analysis of the Ly$\alpha$ transmission at $z>6$ as a function UV magnitude shows a high Ly$\alpha$ transmission of the IGM in UV bright galaxies ($M_{\text{UV}}<-21$), while Ly$\alpha$ visibility from fainter galaxies ($-21<M_{\text{UV}}<-19$) is rapidly suppressed at $z>6$.  This is in general consistent with a picture of the spatially inhomogeneous process of reionization where bright galaxies often created large ionized bubbles which allow enhanced Ly$\alpha$ transmission to the IGM nearby these bright sources, which is illustrared in Figure \ref{fig:cartoon}. 

This picture is consistent with theoretical predictions from reionization simulations where reionization begins in overdense regions and progresses into low-density regions \citep[e.g.,][]{Finlator2009a, Katz2019a}.  In particular, \cite{Park2021a} calculate Ly$\alpha$ transmissivity of the IGM from the Cosmic Dawn II simulation \citep{Ocvirk2020a}, showing that UV-bright galaxies tend to reside in ``bubbles'' as they ionize large volumes and are located in overdense regions where nearby fainter galaxies also can contribute to the ionizing emissivity \citep[][Larson et al. 2021 in preparation]{Finkelstein2019b, Endsley2021b}. These fainter galaxies sit in large ionized bubbles are expected to have higher Ly$\alpha$ transmission than isolated faint objects \citep{Qin2021a}. Most of these faint objects are likely under current detection limit, but could be probed in larger samples (which will be available from Nancy Grace Roman Space Telescope, NGRST) or tested in future observations sensitive to fainter emission from James Webb Space Telescope (JWST) {and} the next generation of 20-30~m-class telescopes. This model also naturally predicts that Ly$\alpha$ emission shows a larger velocity offset from systemic for brighter galaxies, which further facilitates the escape of Ly$\alpha$ photons through the IGM \citep{Mason2018b, Garel2021a}.  \deleted{Therefore,} The existing data from Ly$\alpha$ supports this picture.

Our results suggest significantly different evolution of the Ly$\alpha$ transmission in the IGM in the middle phase of reionization in varying UV-magnitude galaxies.  As the Ly$\alpha$ transmission to the IGM is the key quantity in constraining the neutral fraction of the IGM with Ly$\alpha$ observations \citep{Mason2018a, Hoag2019a, Jung2020a, Morales2021a, Wold2021a}, it is critical to understand the dependence of IGM transmission of Ly$\alpha$ on galaxy UV luminosity in the future. This requires deeper spectroscopic observations and larger sample sizes, both of which will need to cover a wide dynamic range in galaxy UV luminosity. This will be possible by grism observations with NGRST, which will be ideal for such studies of Ly$\alpha$ emission from galaxies during the epoch of reionization. Furthermore, the pending launch of JWST will provide data we can use to study Ly$\alpha$ systematics (e.g., velocity offset) with other emission line observations for galaxies in the reionization era.

\acknowledgments
We thank our colleagues on the CLEAR team for their valuable conversations and contributions.  This work is based on data obtained from the Hubble Space Telescope through program number GO-14227.  Support for Program number GO-14227 was provided by NASA through a grant from the Space Telescope Science Institute, which is operated by the Association of Universities for Research in Astronomy, Incorporated, under NASA contract NAS5-26555. I.J. acknowledges support from NASA under award number 80GSFC21M0002. 
SLF acknowledges support from NASA through ADAP award 80NSSC18K0954.
VEC acknowledges support from the NASA Headquarters under the Future Investigators in NASA Earth and Space Science and Technology (FINESST) award 19-ASTRO19-0122, as well as support from the Hagler Institute for Advanced Study at Texas A\&M University.   This work benefited from generous
support from the George P. and Cynthia Woods Mitchell Institute for
Fundamental Physics and Astronomy at Texas A\&M University.   This work was supported in part by NASA contract  NNG16PJ33C, the Studying Cosmic Dawn with WFIRST Science Investigation Team.  

\appendix

\section{Ly$\alpha$-Emitter Candidate at \lowercase{\textit{z}} = 6.513}
We performed emission-line search as described in Section 2.2.1, which results in no convincing Ly$\alpha$ detection{s} from the entire sample except for one tentative detection from a {$z_{\text{grism}} = 6.513\pm0.005$} galaxy (GN4\_5461276). Although \deleted{we} the detection of Ly$\alpha$ emission is {possibly spurious} from the galaxy with given data, it is worth presenting the promising features of the detected emission line here as the physical properties of the emission line would be rather extreme if confirmed.  Further observations are required for validating the nature of the detected emission line.

\begin{figure*}[ht]
\centering
\includegraphics[width=0.8\paperwidth]{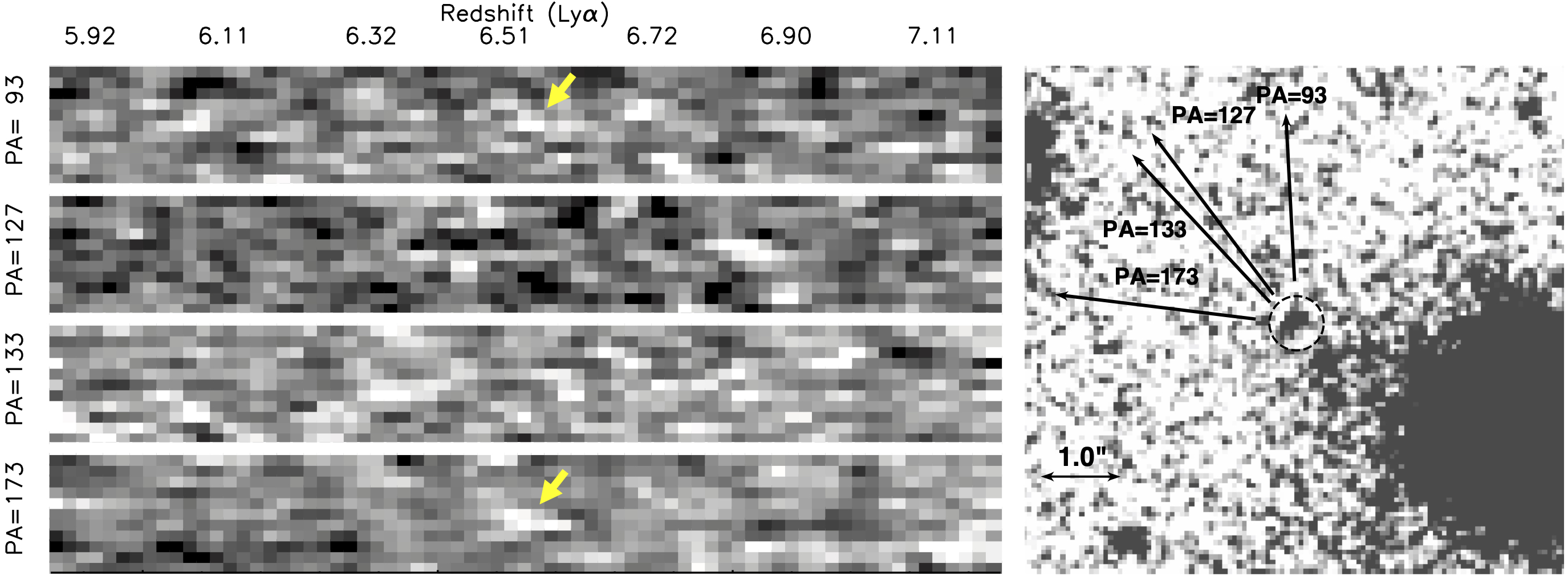}
\caption{(Left) Four individual PA 2D spectra of the emission-line candidate galaxy (GN4\_5461276).  White (black) shows regions with positive (negative) flux in the spectra.  The topmost and bottom-most panels (PA = 93 and 173) show positive emission features.  These are marked with yellow arrows.  The middle two panels (PA=133 and 127) show no such emission features. (Right) HST $H$-band image centered at GN4\_5461276 showing the spectral dispersion directions of the four PAs (as labeled). The two PAs (93 and 173) showing positive emission do not share any potential contaminating sources.  The two intermediate PAs (127 and 133) have possible contamination from the large galaxy in the bottom right of the image.  }
\label{fig:PAs}
\end{figure*}

\begin{figure}[t]
\centering
\includegraphics[width=0.5\columnwidth]{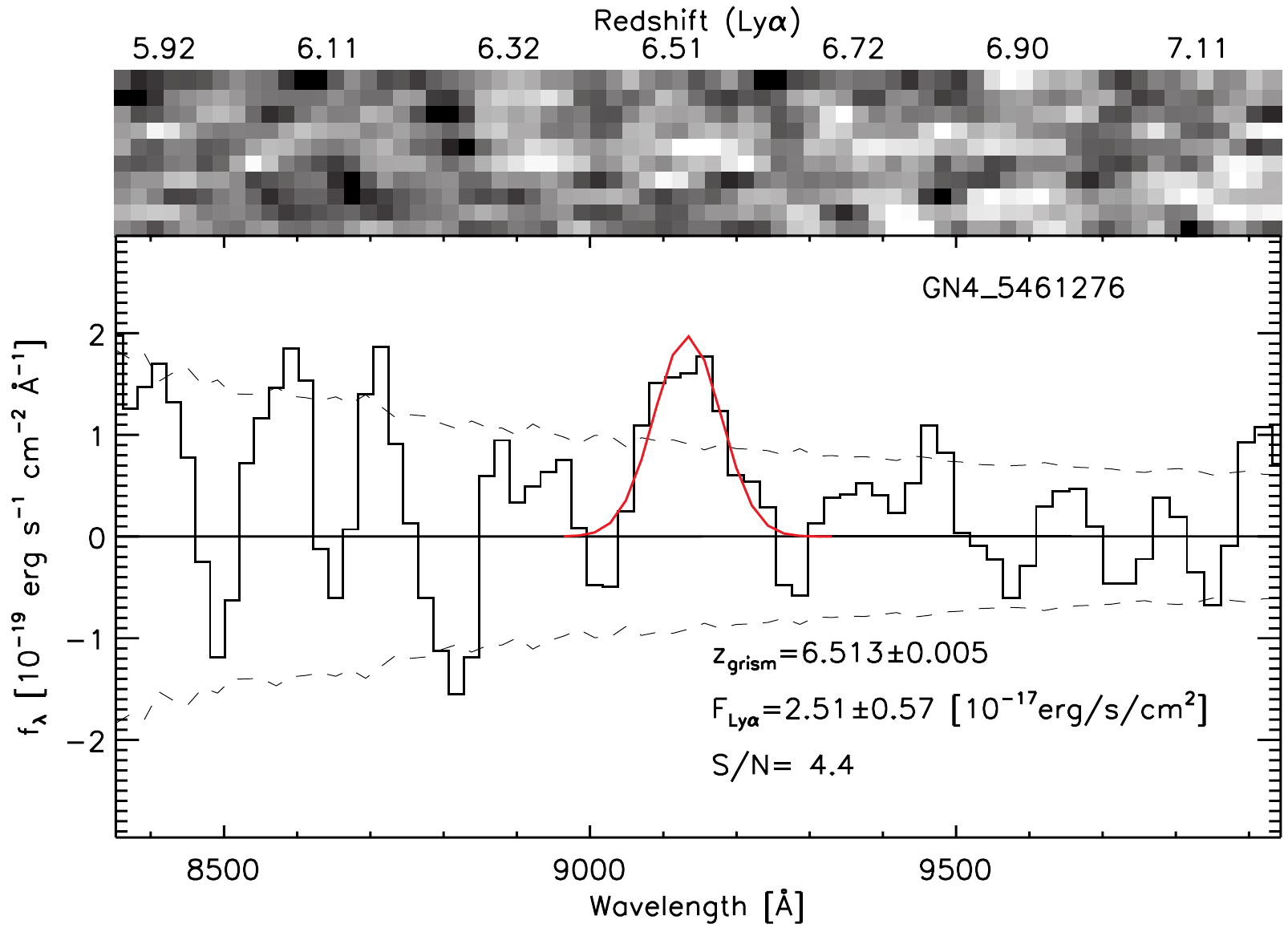}
\caption{All PA combined 1D (bottom) and 2D (top) spectrum of the emission line detected from GN4\_5461276. In the bottom panel, the measured spectrum is shown as the black histogram with the noise level showed as the dashed curves. The red curve represents the a model Gaussian curve fit to the the  emission line.}
\label{fig:lae_spectra}
\end{figure}

We display the individual 2D spectra obtained from 4 PAs with the HST $H$-band stamp of GN4\_5461276 in Figure \ref{fig:PAs} where we detect significant emission-line features from 2 (PA=93 and PA=173) of the 4 PAs. The all PA combined 1D and 2D spectra are shown in Figure \ref{fig:lae_spectra}, which shows a formal $>$4$\sigma$ detection of the emission line.

To rule out the possibility of being contaminated emission from nearby sources, we checked the known redshifts of the nearby bright sources and concluded that the emission line is not {the zeroth-order reflection of an emission line} associated with the nearby sources. We fit the Gaussian function to the emission line to estimate the physical properties of the detected emission line.  The S/N value was estimated from Gaussian fitting on 1000 simulated spectra by fluctuating the 1D fluxes with the 1D noise level.  The emission line flux was calculated from the best-fit Gaussian (red in Figure \ref{fig:lae_spectra}), and the redshift was derived from the peak of the best-fit Gaussian curve. The derived physical quantities are listed in the bottom panel of the plot, which shows its extreme values of a rest-frame EW ($>$1000\AA) and a line width (FWHM$>$100\AA).

\begin{figure*}[t]
\centering
\includegraphics[width=0.8\paperwidth]{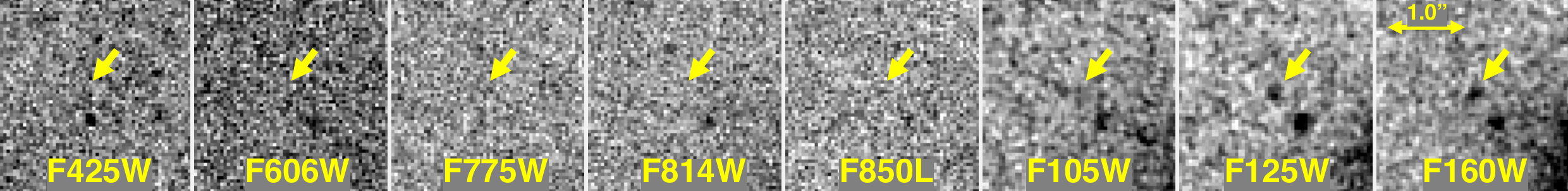}
\includegraphics[width=0.65\paperwidth]{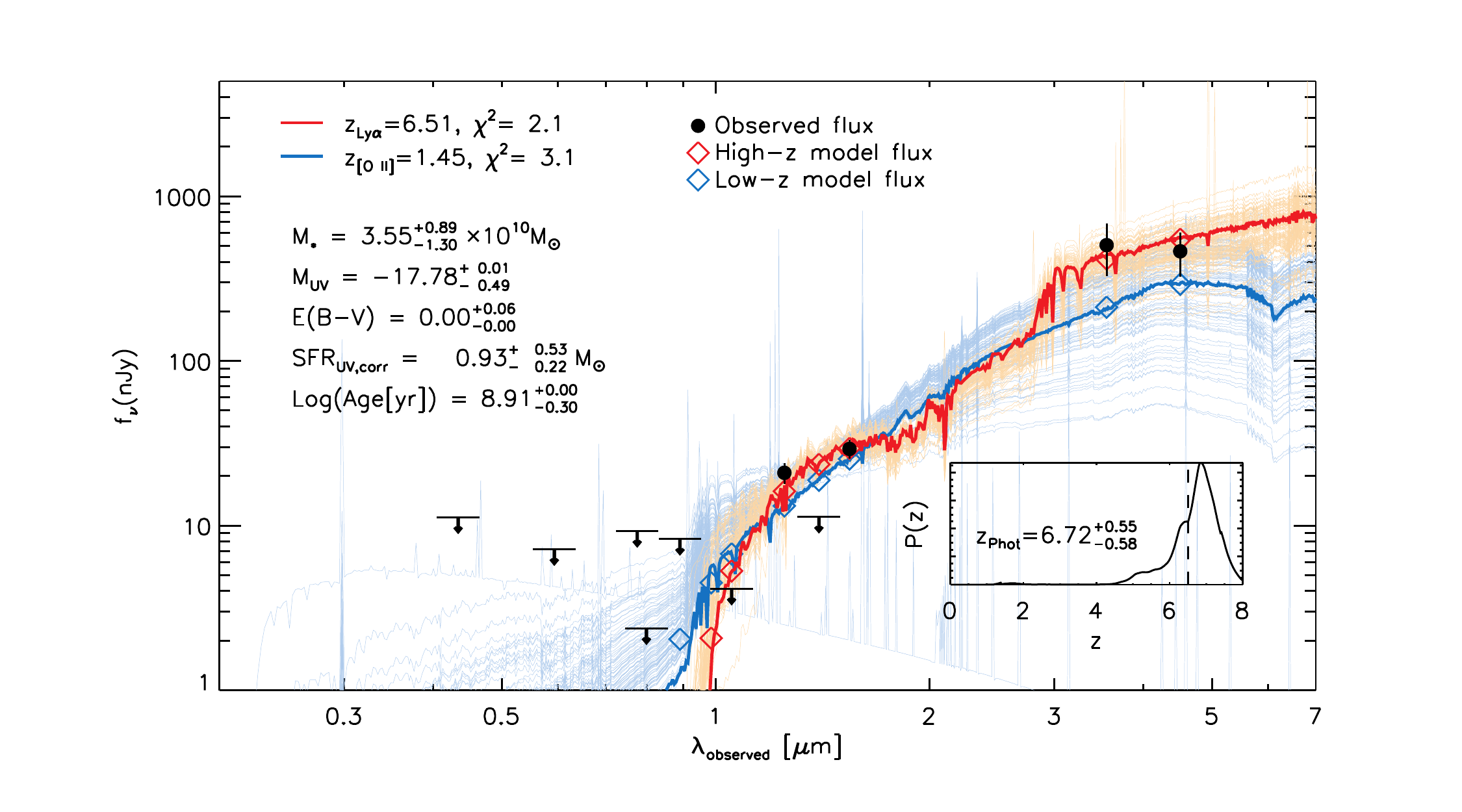}
\caption{Images and SED models for GN4\_5461276.   (Top) The HST ACS/WFC3 ``postage-stamp images'' for this galaxy. The yellow arrows indicate the locations of the object in individual stamps. (Bottom) The best-fit SED models between high-$z$ (red: Ly$\alpha$) and low-$z$ (blue: [\ion{O}{2}]). The faint blue and red curves represent the SED fitting for each 100 MC realization set of the high-$z$ and low-$z$ cases. For the high-$z$ solution, Ly$\alpha$ contribution is subtracted from HST photometry. The inset figure shows the probability distribution function of the photometric redshift (solid curve) with the grism redshift (vertical dashed line). Although the high-$z$ solution ($\chi^2=1.7$) is slightly preferred compared to the low-$z$ ($\chi^2=3.1$), it is difficult to draw conclusion with given data.}
\label{fig:lae_hst}
\end{figure*}

To check the possibility of being a low-$z$ interloper, we compare the best-fit SED models between high-$z$ (Ly$\alpha$) and low-$z$ ([\ion{O}{2}]) interpretations with the HST stamp images in Figure \ref{fig:lae_hst}.  For the high-$z$ interpretation, the Ly$\alpha$ contribution to HST fluxes were subtracted for the SED fitting.  From the best-fit models, the high-$z$ solution is preferred than the low-z solution, but both interpretations remain viable with the given data.  Thus, further observational evidence is required to confirm this tentative Ly$\alpha$ detection. 

\section{Continuum Detection from a \lowercase{\textit{z}} = 6.2 Galaxy}

\begin{figure}[ht]
\centering
\includegraphics[width=0.75\columnwidth]{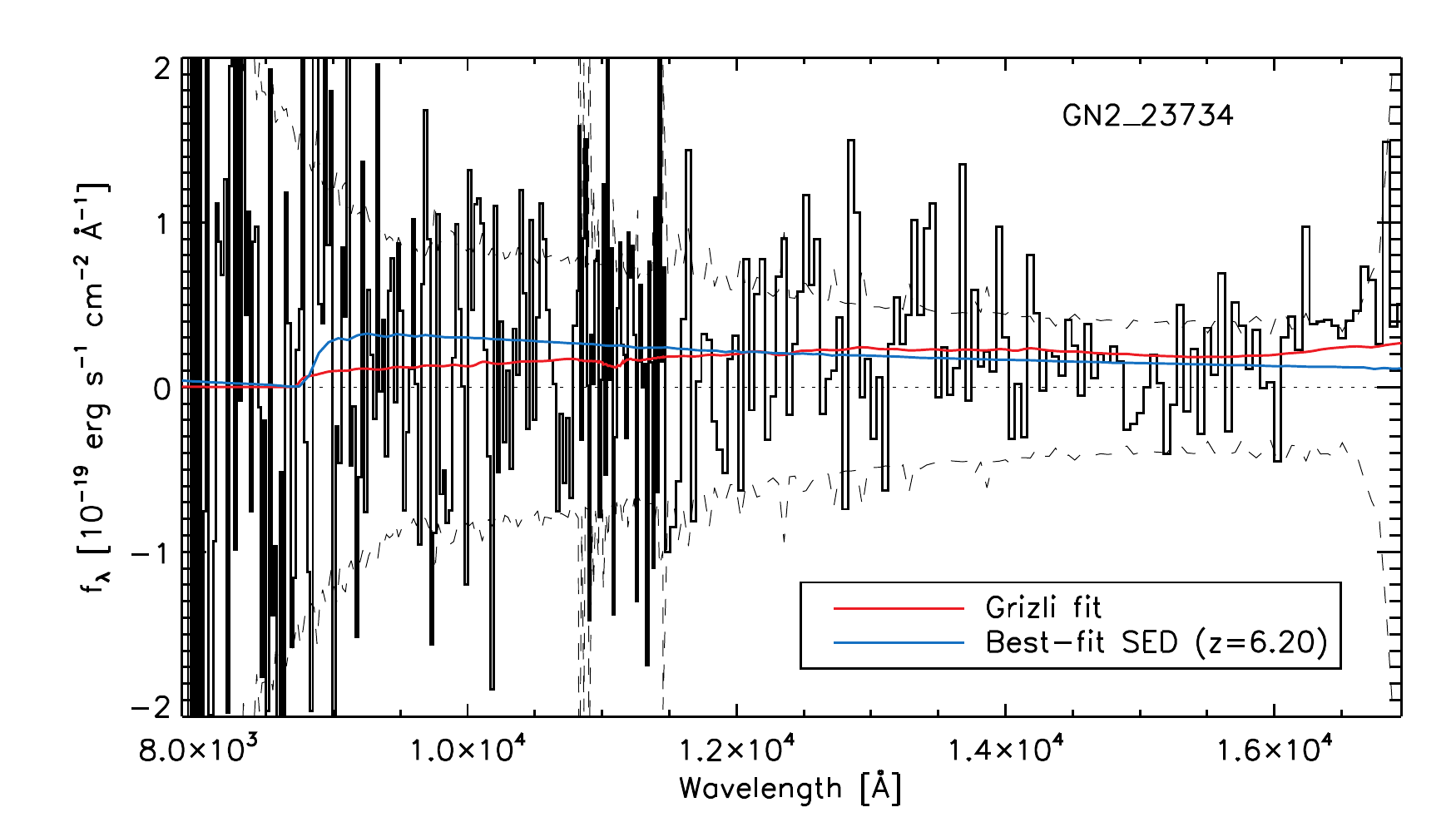}
\caption{The 1D grism spectrum of a continuum-detection candidate galaxy (GN2\_23734) at {$z_{\text{grism}} = 6.2^{+2.5}_{-0.8}$}.  The continuum is formally detected at wavelengths redward of the Ly$\alpha$-break (S/N$_{\text{continuum}}>4$). The black histogram shows the 1D spectrum, which combines both of the G102 and G140 spectra of GN2\_23734. The noise level is shown as dashed curves. The red curve represents the continuum fit derived by {\sc Grizli}, and the best-fit SED is displayed as the blue curve, which is obtained from SED fitting based on the object's HST+Spitzer/IRAC photometry.}
\label{fig:continuum}
\end{figure}

We inspected CLEAR spectra of high-$z$ candidates to see if they present any continuum features as described in Section 2.2.2.  It results in one continuum candidate from a {$z_{\text{grism}} = 6.2^{+2.5}_{-0.8}$} galaxy.  The continuum is detected at $>$4$\sigma$ at wavelengths redward of a possible  continuum break.  Figure \ref{fig:continuum} shows the grism spectra of GN2\_23734, combining both \deleted{of the} G102 and G140 spectra. The extracted 1D spectrum is shown as a black histogram with the noise level shown as dashed-line curves. The continuum fit given by {\sc Grizli} is presented as a red curve, and we display the best-fit SED (blue), which is obtained from the SED fitting based on the object's HST+Spitzer/IRAC photometry (see Section 2.4). In the SED fitting, we used the best-fit grism redshift ($z_{\text{grism}}=6.2$) derived by {\sc Grizli}.  Although the grism continuum-fit spectrum seems comparable to the best-fit SED, we consider it tentative given the current noise level in the grism spectrum.  Nevertheless, we report this tentative detection of the continuum break as it provides a possible path to confirm other high-$z$ galaxies without Ly$\alpha$.  This is particularly useful for galaxies in the epoch of reionization where the Ly$\alpha$ transmission is lower (see Section~2.2.2 and discussion in, e.g., \citealt{Rhoads2013a, Watson2015a, Oesch2016a}).  It will be valuable to target this galaxy (and others) with spectroscopy from future space-based telescopes (e.g., JWST, NGRST).

\section{Summary of $6.0<z<8.2$ Candidate Galaxies in CLEAR}
As discussed in Section 2, we use the CLEAR spectra of 148 $6.0<z<8.2$ candidate galaxies for exploring the galaxy UV magnitude dependency of the IGM transmission to Ly$\alpha$ during the epoch of reionization.  Table \ref{tab:CLEARall} summarizes the entire 148 $6.0<z<8.2$ CLEAR galaxies in order of increasing photometric redshift, which includes the 3$\sigma$ rest-EW upper limits on Ly$\alpha$ in the last column. 
\startlongtable
\begin{deluxetable*}{ccccccc}
\tablecaption{Summary of CLEAR $6.0<z<8.2$ Targets\label{tab:CLEARall}}
\tablehead{ \colhead{ID\tablenotemark{a}} & \colhead{R.A. (J2000.0)}         & \colhead{Decl. (J2000.0)}         & \colhead{$J_{\text{125}}$} & \colhead{$M_{\text{UV} }$\tablenotemark{b}} & \colhead{$z_{\text{phot}}$\tablenotemark{c}} & \colhead{EW$_{\text{Ly}\alpha}$\tablenotemark{d} (\AA)} }
\startdata
{GS4\_28545} & { 53.142085} & {-27.779851} & { 26.12$^{+0.01}_{-0.01}$} & {-20.63$^{+0.02}_{-0.03}$}& {6.00$^{+0.02}_{-0.02}$} & {$<$82.2}\\
{GS4\_5428651} & { 53.162267} & {-27.796050} & { 29.90$^{+0.17}_{-0.15}$} & {-16.03$^{+0.21}_{-0.13}$}& {6.00$^{+5.33}_{-0.85}$} & {$<$1332.7}\\
{GS4\_30518} & { 53.159459} & {-27.771855} & { 28.59$^{+0.06}_{-0.06}$} & {-18.29$^{+0.05}_{-0.02}$}& {6.01$^{+0.08}_{-0.08}$} & {$<$142.8}\\
{GN5\_5453006} & {189.147675} & { 62.310692} & { 26.18$^{+0.10}_{-0.09}$} & {-21.41$^{+0.03}_{-0.09}$}& {6.02$^{+0.09}_{-0.08}$} & {$<$14.0}\\
{GS4\_25097} & { 53.143880} & {-27.793083} & { 27.85$^{+0.02}_{-0.02}$} & {-18.97$^{+0.01}_{-0.02}$}& {6.03$^{+0.07}_{-0.07}$} & {$<$123.9}\\
{GN1\_38168} & {189.320618} & { 62.377956} & { 25.29$^{+0.29}_{-0.23}$} & {-22.97$^{+0.08}_{-0.03}$}& {6.03$^{+0.04}_{-0.05}$} & {$<$3.3}\\
{GS2\_49117} & { 53.128395} & {-27.679525} & { 26.10$^{+0.10}_{-0.09}$} & {-20.41$^{+0.08}_{-0.13}$}& {6.05$^{+0.31}_{-0.40}$} & {$<$55.2}\\
{GN2\_22377} & {189.231018} & { 62.252335} & { 26.71$^{+0.07}_{-0.06}$} & {-20.03$^{+0.07}_{-0.07}$}& {6.06$^{+0.10}_{-0.09}$} & {$<$49.0}\\
{GS4\_24418} & { 53.139299} & {-27.795872} & { 26.85$^{+0.15}_{-0.13}$} & {-19.88$^{+0.05}_{-0.05}$}& {6.07$^{+0.09}_{-0.09}$} & {$<$53.9}\\
{ERSPRIME\_43483} & { 53.070813} & {-27.706732} & { 26.07$^{+0.06}_{-0.06}$} & {-20.66$^{+0.05}_{-0.05}$}& {6.07$^{+0.09}_{-0.09}$} & {$<$36.0}\\
{GS4\_29439} & { 53.165805} & {-27.776102} & { 28.55$^{+0.08}_{-0.08}$} & {-18.23$^{+0.05}_{-0.07}$}& {6.09$^{+0.08}_{-0.08}$} & {$<$155.4}\\
{GS4\_29085} & { 53.151986} & {-27.778248} & { 25.44$^{+0.00}_{-0.00}$} & {-21.36$^{+0.00}_{-0.01}$}& {6.10$^{+0.04}_{-0.05}$} & {$<$8.7}\\
{GN1\_37875} & {189.292984} & { 62.366493} & { 25.96$^{+0.26}_{-0.21}$} & {-20.30$^{+0.15}_{-0.17}$}& {6.12$^{+0.24}_{-0.26}$} & {$<$54.7}\\
{GN4\_5463603} & {189.346069} & { 62.255791} & { 27.21$^{+0.09}_{-0.08}$} & {-19.64$^{+0.08}_{-0.07}$}& {6.13$^{+0.25}_{-0.82}$} & {$<$279.9}\\
{GS2\_49517} & { 53.144795} & {-27.676513} & { 26.54$^{+0.17}_{-0.15}$} & {-20.24$^{+0.14}_{-0.18}$}& {6.14$^{+0.27}_{-0.29}$} & {$<$45.9}\\
{GS5\_40046} & { 53.105382} & {-27.723476} & { 25.56$^{+0.07}_{-0.06}$} & {-21.10$^{+0.07}_{-0.09}$}& {6.15$^{+0.19}_{-0.27}$} & {$<$33.0}\\
{GN5\_5453623} & {189.159882} & { 62.307541} & { 25.76$^{+0.05}_{-0.04}$} & {-20.43$^{+0.09}_{-0.06}$}& {6.15$^{+0.27}_{-0.30}$} & {$<$75.6}\\
{GS4\_24191} & { 53.152647} & {-27.797311} & { 26.81$^{+0.00}_{-0.00}$} & {-20.01$^{+0.05}_{-0.00}$}& {6.15$^{+0.04}_{-0.03}$} & {$<$48.8}\\
{GS4\_24855} & { 53.146101} & {-27.794571} & { 26.77$^{+0.00}_{-0.00}$} & {-19.87$^{+0.01}_{-0.02}$}& {6.17$^{+0.06}_{-0.07}$} & {$<$72.0}\\
{GS3\_5443046} & { 53.155751} & {-27.746295} & { 27.31$^{+0.14}_{-0.13}$} & {-19.39$^{+0.09}_{-0.09}$}& {6.17$^{+0.18}_{-0.19}$} & {$<$103.7}\\
{GS2\_48937} & { 53.123742} & {-27.680620} & { 27.48$^{+0.22}_{-0.18}$} & {-19.29$^{+0.10}_{-0.20}$}& {6.17$^{+0.29}_{-0.40}$} & {$<$89.3}\\
{GS3\_35639} & { 53.155782} & {-27.746150} & { 26.61$^{+0.14}_{-0.13}$} & {-20.28$^{+0.09}_{-0.08}$}& {6.17$^{+0.11}_{-0.11}$} & {$<$39.6}\\
{GS5\_44454} & { 53.120956} & {-27.702285} & { 26.19$^{+0.13}_{-0.12}$} & {-20.61$^{+0.14}_{-0.11}$}& {6.18$^{+0.17}_{-0.22}$} & {$<$26.4}\\
{ERSPRIME\_43514} & { 53.071191} & {-27.706498} & { 26.30$^{+0.09}_{-0.08}$} & {-20.40$^{+0.07}_{-0.07}$}& {6.18$^{+0.13}_{-0.13}$} & {$<$45.3}\\
{GN3\_30188} & {189.256210} & { 62.291283} & { 27.02$^{+0.05}_{-0.05}$} & {-19.67$^{+0.06}_{-0.04}$}& {6.20$^{+0.09}_{-0.10}$} & {$<$242.8}\\
{GN7\_5429834} & {189.095474} & { 62.227612} & { 27.77$^{+0.23}_{-0.19}$} & {-19.14$^{+0.17}_{-0.13}$}& {6.20$^{+0.26}_{-0.63}$} & {$<$112.1}\\
{GN3\_5450671} & {189.263977} & { 62.318367} & { 27.64$^{+0.17}_{-0.15}$} & {-19.49$^{+0.08}_{-0.07}$}& {6.21$^{+0.10}_{-0.09}$} & {$<$94.7}\\
{GN4\_5461825} & {189.323929} & { 62.271130} & { 28.16$^{+0.12}_{-0.11}$} & {-18.76$^{+0.09}_{-0.08}$}& {6.22$^{+0.20}_{-0.22}$} & {$<$174.3}\\
{GN4\_5446203} & {189.305573} & { 62.271095} & { 27.96$^{+0.22}_{-0.18}$} & {-17.87$^{+0.88}_{-0.49}$}& {6.22$^{+2.25}_{-4.08}$} & {$<$14672.0}\\
{GN4\_23534} & {189.296127} & { 62.257969} & { 26.88$^{+0.13}_{-0.12}$} & {-19.98$^{+0.14}_{-0.10}$}& {6.23$^{+0.21}_{-0.33}$} & {$<$47.7}\\
{GN2\_5432549} & {189.252869} & { 62.235577} & { 27.63$^{+0.09}_{-0.08}$} & {-19.33$^{+0.06}_{-0.05}$}& {6.24$^{+0.11}_{-0.12}$} & {$<$128.9}\\
{ERSPRIME\_5455619} & { 53.069652} & {-27.697069} & { 25.63$^{+0.02}_{-0.02}$} & {-21.06$^{+0.02}_{-0.02}$}& {6.24$^{+0.13}_{-0.13}$} & {$<$67.6}\\
{GN5\_5453421} & {189.157303} & { 62.309128} & { 27.85$^{+0.17}_{-0.15}$} & {-19.07$^{+0.16}_{-0.11}$}& {6.24$^{+0.33}_{-0.88}$} & {$<$146.2}\\
{GN4\_21677} & {189.318161} & { 62.249264} & { 26.91$^{+0.10}_{-0.09}$} & {-19.95$^{+0.10}_{-0.08}$}& {6.28$^{+0.15}_{-0.15}$} & {$<$61.5}\\
{GN7\_16571} & {189.114838} & { 62.224430} & { 25.73$^{+0.04}_{-0.04}$} & {-21.08$^{+0.05}_{-0.05}$}& {6.30$^{+0.06}_{-0.07}$} & {$<$18.0}\\
{GS5\_5445164} & { 53.114526} & {-27.737678} & { 26.95$^{+0.17}_{-0.15}$} & {-19.81$^{+0.18}_{-0.16}$}& {6.30$^{+0.32}_{-4.74}$} & {$<$87.2}\\
{GN2\_23734} & {189.211990} & { 62.258846} & { 26.24$^{+0.06}_{-0.06}$} & {-20.43$^{+0.04}_{-0.04}$}& {6.31$^{+0.18}_{-0.21}$} & {$<$67.4}\\
{GS4\_29685} & { 53.156131} & {-27.775849} & { 26.42$^{+0.00}_{-0.00}$} & {-20.22$^{+0.01}_{-0.01}$}& {6.32$^{+0.06}_{-0.07}$} & {$<$38.0}\\
{GN2\_17132} & {189.197571} & { 62.226990} & { 26.30$^{+0.04}_{-0.04}$} & {-20.48$^{+0.03}_{-0.03}$}& {6.32$^{+0.08}_{-0.08}$} & {$<$34.7}\\
{GN2\_19472} & {189.219406} & { 62.238247} & { 27.70$^{+0.08}_{-0.08}$} & {-18.97$^{+0.08}_{-0.07}$}& {6.32$^{+0.20}_{-0.21}$} & {$<$174.7}\\
{GS2\_47924} & { 53.146726} & {-27.686787} & { 26.96$^{+0.08}_{-0.07}$} & {-19.86$^{+0.10}_{-0.12}$}& {6.33$^{+0.30}_{-0.29}$} & {$<$68.9}\\
{GS5\_40767} & { 53.118464} & {-27.719486} & { 26.36$^{+0.11}_{-0.10}$} & {-20.43$^{+0.10}_{-0.12}$}& {6.34$^{+0.24}_{-0.23}$} & {$<$35.4}\\
{GS4\_28201} & { 53.156040} & {-27.780970} & { 27.86$^{+0.02}_{-0.02}$} & {-19.04$^{+0.01}_{-0.01}$}& {6.34$^{+0.04}_{-0.56}$} & {$<$71.6}\\
{GN3\_35300} & {189.238388} & { 62.327126} & { 25.51$^{+0.08}_{-0.08}$} & {-21.26$^{+0.12}_{-0.09}$}& {6.35$^{+0.19}_{-0.19}$} & {$<$22.4}\\
{GS4\_23893} & { 53.153257} & {-27.798288} & { 27.69$^{+0.01}_{-0.01}$} & {-18.87$^{+0.02}_{-0.01}$}& {6.38$^{+0.07}_{-0.06}$} & {$<$180.2}\\
{GN1\_37866} & {189.323151} & { 62.366215} & { 25.99$^{+0.16}_{-0.14}$} & {-20.11$^{+0.10}_{-0.13}$}& {6.39$^{+0.57}_{-1.08}$} & {$<$76.4}\\
{GS3\_34282} & { 53.149862} & {-27.752839} & { 27.25$^{+0.29}_{-0.23}$} & {-19.82$^{+0.15}_{-0.12}$}& {6.42$^{+0.24}_{-0.64}$} & {$<$224.2}\\
{GN1\_37461} & {189.318008} & { 62.353180} & { 26.50$^{+0.12}_{-0.11}$} & {-20.22$^{+0.09}_{-0.08}$}& {6.43$^{+0.14}_{-0.30}$} & {$<$65.5}\\
{GS4\_27209} & { 53.173343} & {-27.784645} & { 28.05$^{+0.03}_{-0.03}$} & {-18.90$^{+0.01}_{-0.02}$}& {6.44$^{+0.06}_{-0.04}$} & {$<$88.9}\\
{ERSPRIME\_44510} & { 53.049653} & {-27.701990} & { 26.85$^{+0.05}_{-0.04}$} & {-19.86$^{+0.04}_{-0.07}$}& {6.45$^{+0.07}_{-0.08}$} & {$<$98.9}\\
{GN4\_27400} & {189.277420} & { 62.276405} & { 25.92$^{+0.04}_{-0.04}$} & {-20.95$^{+0.04}_{-0.04}$}& {6.45$^{+0.09}_{-0.10}$} & {$<$26.0}\\
{GS2\_5457820} & { 53.114407} & {-27.685098} & { 26.91$^{+0.16}_{-0.14}$} & {-19.31$^{+0.09}_{-0.13}$}& {6.47$^{+0.47}_{-0.49}$} & {$<$360.3}\\
{GS4\_5430868} & { 53.155602} & {-27.788733} & { 29.47$^{+0.05}_{-0.05}$} & {-17.48$^{+0.08}_{-0.03}$}& {6.48$^{+0.13}_{-0.12}$} & {$<$355.1}\\
{GN2\_20362} & {189.240814} & { 62.242630} & { 26.91$^{+0.06}_{-0.05}$} & {-20.04$^{+0.05}_{-0.04}$}& {6.52$^{+0.08}_{-0.08}$} & {$<$53.5}\\
{GS3\_40377} & { 53.172480} & {-27.721434} & { 26.42$^{+0.08}_{-0.07}$} & {-20.52$^{+0.05}_{-0.07}$}& {6.54$^{+0.13}_{-0.14}$} & {$<$38.0}\\
{ERSPRIME\_39697} & { 53.064237} & {-27.724698} & { 27.11$^{+0.10}_{-0.09}$} & {-19.77$^{+0.08}_{-0.10}$}& {6.55$^{+0.14}_{-0.15}$} & {$<$83.6}\\
{GN5\_33256} & {189.178146} & { 62.310635} & { 27.41$^{+0.06}_{-0.06}$} & {-19.36$^{+0.08}_{-0.06}$}& {6.57$^{+0.31}_{-0.33}$} & {$<$128.3}\\
{GN7\_16422} & {189.177979} & { 62.223713} & { 26.32$^{+0.05}_{-0.05}$} & {-20.49$^{+0.07}_{-0.05}$}& {6.58$^{+0.11}_{-0.12}$} & {$<$71.4}\\
{GN7\_14851} & {189.111633} & { 62.215374} & { 26.54$^{+0.03}_{-0.03}$} & {-20.33$^{+0.03}_{-0.04}$}& {6.58$^{+0.06}_{-0.07}$} & {$<$43.4}\\
{GN4\_5438687} & {189.295410} & { 62.252560} & { 28.66$^{+0.28}_{-0.22}$} & {-18.25$^{+0.24}_{-0.13}$}& {6.59$^{+0.52}_{-5.20}$} & {$<$330.2}\\
{GS4\_28784} & { 53.169042} & {-27.778832} & { 28.06$^{+0.03}_{-0.03}$} & {-18.79$^{+0.02}_{-0.03}$}& {6.61$^{+0.06}_{-0.06}$} & {$<$129.3}\\
{GN5\_34340} & {189.187347} & { 62.318821} & { 27.27$^{+0.08}_{-0.07}$} & {-19.71$^{+0.06}_{-0.07}$}& {6.63$^{+0.18}_{-0.17}$} & {$<$66.0}\\
{GN2\_5465437} & {189.239700} & { 62.248108} & { 26.35$^{+0.01}_{-0.01}$} & {-20.59$^{+0.01}_{-0.01}$}& {6.65$^{+0.04}_{-0.05}$} & {$<$39.9}\\
{GS4\_5431395} & { 53.161653} & {-27.787043} & { 27.97$^{+0.05}_{-0.05}$} & {-18.88$^{+0.07}_{-0.03}$}& {6.67$^{+0.13}_{-0.14}$} & {$<$128.6}\\
{ERSPRIME\_43078} & { 53.055896} & {-27.708708} & { 26.74$^{+0.09}_{-0.08}$} & {-20.27$^{+0.06}_{-0.06}$}& {6.68$^{+0.12}_{-0.13}$} & {$<$65.6}\\
{GS3\_5447463} & { 53.167998} & {-27.728187} & { 28.31$^{+0.24}_{-0.20}$} & {-18.34$^{+0.17}_{-0.14}$}& {6.70$^{+0.63}_{-0.79}$} & {$<$689.3}\\
{GN4\_5462173} & {189.308792} & { 62.267353} & { 26.66$^{+0.05}_{-0.05}$} & {-20.12$^{+0.04}_{-0.06}$}& {6.71$^{+0.25}_{-0.34}$} & {$<$125.9}\\
{GN7\_5430295} & {189.088989} & { 62.229202} & { 28.07$^{+0.15}_{-0.13}$} & {-18.96$^{+0.11}_{-0.09}$}& {6.72$^{+0.25}_{-0.26}$} & {$<$278.5}\\
{GN5\_31683} & {189.165039} & { 62.300194} & { 26.35$^{+0.01}_{-0.01}$} & {-20.59$^{+0.04}_{-0.02}$}& {6.73$^{+0.05}_{-0.05}$} & {$<$35.5}\\
{GN5\_32053} & {189.140656} & { 62.302368} & { 27.40$^{+0.15}_{-0.13}$} & {-19.47$^{+0.09}_{-0.14}$}& {6.73$^{+0.46}_{-0.57}$} & {$<$157.3}\\
{GN5\_33584} & {189.156357} & { 62.313087} & { 27.11$^{+0.08}_{-0.07}$} & {-19.67$^{+0.08}_{-0.08}$}& {6.75$^{+0.50}_{-0.45}$} & {$<$112.4}\\
{GN7\_5428106} & {189.104630} & { 62.222271} & { 27.49$^{+0.14}_{-0.12}$} & {-19.59$^{+0.10}_{-0.08}$}& {6.75$^{+0.18}_{-0.20}$} & {$<$69.1}\\
{GS4\_26311} & { 53.151601} & {-27.787910} & { 28.07$^{+0.04}_{-0.03}$} & {-18.91$^{+0.02}_{-0.03}$}& {6.76$^{+0.07}_{-0.07}$} & {$<$123.5}\\
{GN5\_5454319} & {189.140610} & { 62.305511} & { 26.14$^{+0.17}_{-0.14}$} & {-20.73$^{+0.10}_{-0.13}$}& {6.77$^{+0.32}_{-0.40}$} & {$<$46.7}\\
{GS4\_20530} & { 53.154950} & {-27.815805} & { 25.89$^{+0.08}_{-0.07}$} & {-20.86$^{+0.04}_{-0.11}$}& {6.79$^{+0.07}_{-0.07}$} & {$<$35.6}\\
{GN4\_5461276\tablenotemark{e}} & {189.281143} & { 62.274857} & { 28.10$^{+0.17}_{-0.15}$} & {-18.01$^{+0.24}_{-0.26}$}& {6.80$^{+0.56}_{-0.60}$} & {-}\\
{ERSPRIME\_45005} & { 53.066739} & {-27.699829} & { 26.09$^{+0.18}_{-0.16}$} & {-20.86$^{+0.13}_{-0.17}$}& {6.81$^{+0.30}_{-0.49}$} & {$<$29.6}\\
{GN4\_26634} & {189.282623} & { 62.272507} & { 27.01$^{+0.08}_{-0.07}$} & {-19.90$^{+0.05}_{-0.09}$}& {6.82$^{+0.24}_{-0.26}$} & {$<$89.3}\\
{GS5\_41330} & { 53.094447} & {-27.716946} & { 25.53$^{+0.10}_{-0.09}$} & {-21.37$^{+0.09}_{-0.10}$}& {6.82$^{+0.16}_{-0.16}$} & {$<$16.8}\\
{ERSPRIME\_39962} & { 53.041104} & {-27.723434} & { 26.60$^{+0.14}_{-0.12}$} & {-20.43$^{+0.10}_{-0.08}$}& {6.83$^{+0.12}_{-0.16}$} & {$<$55.2}\\
{GN3\_5452837} & {189.220932} & { 62.311447} & { 27.84$^{+0.12}_{-0.11}$} & {-19.16$^{+0.06}_{-0.11}$}& {6.83$^{+0.25}_{-0.25}$} & {$<$154.1}\\
{GN5\_34059} & {189.166245} & { 62.316494} & { 26.36$^{+0.05}_{-0.05}$} & {-20.57$^{+0.04}_{-0.05}$}& {6.87$^{+0.15}_{-0.14}$} & {$<$40.7}\\
{GS3\_35821} & { 53.150039} & {-27.745016} & { 26.18$^{+0.10}_{-0.09}$} & {-20.77$^{+0.07}_{-0.08}$}& {6.88$^{+0.18}_{-0.21}$} & {$<$27.4}\\
{GS4\_20622} & { 53.155370} & {-27.815248} & { 25.01$^{+0.03}_{-0.03}$} & {-21.88$^{+0.01}_{-0.03}$}& {6.88$^{+0.04}_{-0.03}$} & {$<$14.0}\\
{GN2\_5430417} & {189.223526} & { 62.229626} & { 27.77$^{+0.14}_{-0.12}$} & {-19.25$^{+0.09}_{-0.10}$}& {6.89$^{+0.20}_{-0.22}$} & {$<$126.7}\\
{GN5\_32855} & {189.134445} & { 62.307865} & { 26.36$^{+0.06}_{-0.06}$} & {-20.52$^{+0.05}_{-0.07}$}& {6.90$^{+0.21}_{-0.19}$} & {$<$57.7}\\
{GS4\_25335} & { 53.177375} & {-27.792132} & { 27.17$^{+0.03}_{-0.03}$} & {-19.70$^{+0.02}_{-0.02}$}& {6.90$^{+0.09}_{-0.09}$} & {$<$77.8}\\
{GN7\_15746} & {189.095871} & { 62.220078} & { 26.14$^{+0.03}_{-0.03}$} & {-20.72$^{+0.03}_{-0.03}$}& {6.91$^{+0.16}_{-0.15}$} & {$<$42.0}\\
{GS2\_5452228} & { 53.136772} & {-27.710775} & { 26.41$^{+0.08}_{-0.08}$} & {-20.48$^{+0.07}_{-0.08}$}& {6.91$^{+0.15}_{-0.15}$} & {$<$48.0}\\
{GS4\_22623} & { 53.158407} & {-27.804372} & { 28.69$^{+0.08}_{-0.07}$} & {-18.04$^{+0.06}_{-0.03}$}& {6.91$^{+0.21}_{-0.20}$} & {$<$330.6}\\
{GN4\_24681} & {189.353012} & { 62.263409} & { 25.66$^{+0.02}_{-0.02}$} & {-21.40$^{+0.02}_{-0.02}$}& {6.94$^{+0.05}_{-0.06}$} & {$<$43.0}\\
{GN5\_35003} & {189.178085} & { 62.324314} & { 26.84$^{+0.06}_{-0.05}$} & {-20.15$^{+0.06}_{-0.05}$}& {6.94$^{+0.16}_{-0.16}$} & {$<$68.1}\\
{GN3\_5455960} & {189.279175} & { 62.299282} & { 27.40$^{+0.10}_{-0.09}$} & {-19.64$^{+0.10}_{-0.09}$}& {6.95$^{+0.26}_{-0.30}$} & {$<$173.9}\\
{GS4\_23998} & { 53.160569} & {-27.797819} & { 28.99$^{+0.04}_{-0.04}$} & {-17.73$^{+0.02}_{-0.01}$}& {6.97$^{+0.07}_{-0.08}$} & {$<$306.7}\\
{GN3\_5455082} & {189.229721} & { 62.302711} & { 28.16$^{+0.11}_{-0.10}$} & {-18.91$^{+0.07}_{-0.08}$}& {6.97$^{+0.31}_{-0.37}$} & {$<$291.4}\\
{GN7\_11318} & {189.116867} & { 62.198677} & { 26.66$^{+0.25}_{-0.20}$} & {-20.23$^{+0.21}_{-0.15}$}& {6.98$^{+0.36}_{-0.37}$} & {$<$175.6}\\
{ERSPRIME\_38389} & { 53.066731} & {-27.731170} & { 27.51$^{+0.08}_{-0.07}$} & {-19.48$^{+0.06}_{-0.05}$}& {6.98$^{+0.21}_{-0.28}$} & {$<$343.1}\\
{GN5\_33361} & {189.177856} & { 62.311707} & { 26.40$^{+0.06}_{-0.06}$} & {-20.59$^{+0.04}_{-0.05}$}& {6.99$^{+0.20}_{-0.21}$} & {$<$35.5}\\
{GN4\_26575} & {189.283737} & { 62.272240} & { 26.90$^{+0.04}_{-0.04}$} & {-19.89$^{+0.04}_{-0.09}$}& {6.99$^{+0.31}_{-0.31}$} & {$<$166.5}\\
{GN2\_24019} & {189.275925} & { 62.260296} & { 27.14$^{+0.05}_{-0.05}$} & {-19.85$^{+0.04}_{-0.04}$}& {7.01$^{+0.18}_{-0.18}$} & {$<$262.4}\\
{GN2\_17220} & {189.201050} & { 62.227440} & { 27.08$^{+0.07}_{-0.07}$} & {-19.87$^{+0.06}_{-0.04}$}& {7.02$^{+0.21}_{-0.21}$} & {$<$77.4}\\
{GN7\_5428506} & {189.150253} & { 62.223709} & { 28.01$^{+0.22}_{-0.19}$} & {-19.26$^{+0.15}_{-0.07}$}& {7.03$^{+0.27}_{-0.43}$} & {$<$133.2}\\
{GN5\_34042} & {189.166092} & { 62.316406} & { 27.75$^{+0.17}_{-0.15}$} & {-19.13$^{+0.14}_{-0.20}$}& {7.03$^{+0.60}_{-5.54}$} & {$<$232.7}\\
{GN4\_5441249} & {189.310867} & { 62.260452} & { 27.20$^{+0.03}_{-0.03}$} & {-19.76$^{+0.03}_{-0.02}$}& {7.04$^{+0.14}_{-0.13}$} & {$<$98.6}\\
{GN1\_37724} & {189.273300} & { 62.360783} & { 25.80$^{+0.18}_{-0.15}$} & {-21.18$^{+0.12}_{-0.08}$}& {7.05$^{+0.21}_{-0.21}$} & {$<$25.4}\\
{GN5\_31436} & {189.166931} & { 62.298553} & { 28.24$^{+0.17}_{-0.15}$} & {-19.03$^{+0.10}_{-0.10}$}& {7.06$^{+0.25}_{-5.53}$} & {$<$261.7}\\
{GN3\_34711} & {189.254974} & { 62.321877} & { 27.16$^{+0.20}_{-0.17}$} & {-19.82$^{+0.14}_{-0.17}$}& {7.06$^{+0.34}_{-0.34}$} & {$<$113.6}\\
{GN2\_5464240} & {189.232895} & { 62.247395} & { 28.68$^{+0.20}_{-0.17}$} & {-18.45$^{+0.08}_{-0.12}$}& {7.06$^{+0.36}_{-5.52}$} & {$<$399.1}\\
{GN2\_20956} & {189.205231} & { 62.245541} & { 27.64$^{+0.08}_{-0.07}$} & {-19.20$^{+0.04}_{-0.08}$}& {7.07$^{+0.31}_{-0.32}$} & {$<$213.9}\\
{GN7\_12243} & {189.114151} & { 62.203175} & { 25.47$^{+0.05}_{-0.05}$} & {-21.49$^{+0.08}_{-0.07}$}& {7.08$^{+0.17}_{-0.20}$} & {$<$26.2}\\
{GS4\_5435288} & { 53.153420} & {-27.774471} & { 27.30$^{+0.22}_{-0.18}$} & {-19.62$^{+0.06}_{-0.18}$}& {7.11$^{+0.33}_{-5.53}$} & {$<$111.9}\\
{GN7\_16103} & {189.091812} & { 62.221901} & { 27.32$^{+0.06}_{-0.06}$} & {-19.73$^{+0.05}_{-0.06}$}& {7.14$^{+0.26}_{-0.25}$} & {$<$103.2}\\
{GN4\_5463508} & {189.307037} & { 62.255745} & { 28.26$^{+0.09}_{-0.09}$} & {-18.82$^{+0.06}_{-0.08}$}& {7.18$^{+0.29}_{-0.27}$} & {$<$282.2}\\
{GN3\_5449006} & {189.286926} & { 62.277760} & { 27.42$^{+0.06}_{-0.05}$} & {-19.59$^{+0.04}_{-0.05}$}& {7.19$^{+0.17}_{-0.18}$} & {$<$299.1}\\
{GN3\_5454301} & {189.225098} & { 62.306057} & { 28.58$^{+0.31}_{-0.24}$} & {-18.51$^{+0.17}_{-0.21}$}& {7.19$^{+0.53}_{-5.35}$} & {$<$442.2}\\
{GN2\_21995} & {189.205307} & { 62.250763} & { 26.44$^{+0.06}_{-0.06}$} & {-20.59$^{+0.05}_{-0.04}$}& {7.22$^{+0.18}_{-0.19}$} & {$<$39.5}\\
{GN4\_25192} & {189.264938} & { 62.265793} & { 26.06$^{+0.02}_{-0.02}$} & {-20.95$^{+0.02}_{-0.02}$}& {7.22$^{+0.14}_{-0.14}$} & {$<$41.5}\\
{GN1\_37619} & {189.278732} & { 62.357456} & { 26.21$^{+0.16}_{-0.14}$} & {-20.99$^{+0.08}_{-0.06}$}& {7.23$^{+0.13}_{-0.13}$} & {$<$37.7}\\
{GN7\_5429974} & {189.091492} & { 62.228096} & { 27.87$^{+0.12}_{-0.11}$} & {-19.13$^{+0.10}_{-0.09}$}& {7.28$^{+0.61}_{-0.98}$} & {$<$305.1}\\
{GS4\_5427001} & { 53.152865} & {-27.801940} & { 29.31$^{+0.16}_{-0.14}$} & {-17.40$^{+0.09}_{-0.09}$}& {7.30$^{+0.52}_{-0.57}$} & {$<$848.5}\\
{GS3\_37079} & { 53.169588} & {-27.738066} & { 26.76$^{+0.16}_{-0.14}$} & {-20.30$^{+0.11}_{-0.11}$}& {7.32$^{+0.32}_{-0.35}$} & {$<$55.0}\\
{GN3\_35055} & {189.273392} & { 62.324780} & { 26.15$^{+0.31}_{-0.24}$} & {-20.90$^{+0.17}_{-0.13}$}& {7.35$^{+0.29}_{-0.30}$} & {$<$36.0}\\
{GN5\_32031} & {189.157898} & { 62.302376} & { 25.54$^{+0.02}_{-0.02}$} & {-21.52$^{+0.02}_{-0.02}$}& {7.37$^{+0.15}_{-0.15}$} & {$<$27.8}\\
{GN2\_21790} & {189.199692} & { 62.249802} & { 27.55$^{+0.06}_{-0.06}$} & {-19.48$^{+0.05}_{-0.04}$}& {7.37$^{+0.26}_{-0.25}$} & {$<$113.0}\\
{GN3\_33949} & {189.222000} & { 62.315758} & { 26.98$^{+0.07}_{-0.06}$} & {-19.99$^{+0.06}_{-0.05}$}& {7.39$^{+0.19}_{-0.18}$} & {$<$79.0}\\
{GN5\_33655} & {189.133728} & { 62.313564} & { 26.59$^{+0.10}_{-0.09}$} & {-20.52$^{+0.07}_{-0.07}$}& {7.43$^{+0.21}_{-0.22}$} & {$<$64.3}\\
{GS4\_23143} & { 53.155096} & {-27.801771} & { 27.83$^{+0.01}_{-0.01}$} & {-19.22$^{+0.00}_{-0.00}$}& {7.44$^{+0.14}_{-0.14}$} & {$<$154.4}\\
{GS4\_27055} & { 53.161714} & {-27.785390} & { 26.80$^{+0.03}_{-0.03}$} & {-20.34$^{+0.01}_{-0.02}$}& {7.44$^{+0.10}_{-0.09}$} & {$<$38.4}\\
{GS4\_27958} & { 53.138064} & {-27.781866} & { 29.04$^{+0.05}_{-0.05}$} & {-18.27$^{+0.02}_{-0.02}$}& {7.48$^{+0.23}_{-0.24}$} & {$<$846.2}\\
{GN4\_23416} & {189.333069} & { 62.257233} & { 25.91$^{+0.02}_{-0.02}$} & {-21.19$^{+0.01}_{-0.02}$}& {7.54$^{+0.24}_{-0.26}$} & {$<$66.7}\\
{GS3\_36060} & { 53.172602} & {-27.743931} & { 25.96$^{+0.07}_{-0.06}$} & {-21.11$^{+0.04}_{-0.05}$}& {7.54$^{+0.39}_{-0.41}$} & {$<$43.6}\\
{GN2\_20084} & {189.249817} & { 62.241226} & { 26.32$^{+0.03}_{-0.03}$} & {-20.82$^{+0.07}_{-0.03}$}& {7.58$^{+0.15}_{-0.15}$} & {$<$39.7}\\
{GS4\_5432952} & { 53.170708} & {-27.782452} & { 29.12$^{+0.09}_{-0.08}$} & {-17.68$^{+0.05}_{-0.10}$}& {7.63$^{+0.23}_{-5.84}$} & {$<$916.7}\\
{GN7\_11906} & {189.117416} & { 62.201412} & { 26.47$^{+0.27}_{-0.21}$} & {-20.61$^{+0.23}_{-0.13}$}& {7.65$^{+0.65}_{-0.70}$} & {$<$42.2}\\
{GN7\_16124} & {189.084152} & { 62.222023} & { 25.79$^{+0.03}_{-0.03}$} & {-21.37$^{+0.04}_{-0.03}$}& {7.68$^{+0.37}_{-0.34}$} & {$<$34.2}\\
{GN2\_5436513} & {189.245880} & { 62.244953} & { 28.04$^{+0.17}_{-0.14}$} & {-19.07$^{+0.12}_{-0.11}$}& {7.69$^{+0.36}_{-0.64}$} & {$<$230.4}\\
{GN2\_23331} & {189.197769} & { 62.256966} & { 27.67$^{+0.12}_{-0.10}$} & {-19.52$^{+0.08}_{-0.08}$}& {7.71$^{+0.32}_{-0.45}$} & {$<$274.8}\\
{GN2\_20338} & {189.203125} & { 62.242481} & { 26.48$^{+0.07}_{-0.06}$} & {-20.66$^{+0.04}_{-0.06}$}& {7.72$^{+0.20}_{-0.21}$} & {$<$42.2}\\
{GN5\_36116} & {189.161255} & { 62.334435} & { 25.73$^{+0.05}_{-0.05}$} & {-21.47$^{+0.04}_{-0.04}$}& {7.75$^{+0.54}_{-0.52}$} & {$<$44.5}\\
{GN4\_28143} & {189.300125} & { 62.280354} & { 27.23$^{+0.21}_{-0.18}$} & {-19.90$^{+0.14}_{-0.13}$}& {7.75$^{+0.48}_{-0.78}$} & {$<$99.6}\\
{GN4\_22848} & {189.336029} & { 62.254650} & { 27.26$^{+0.08}_{-0.07}$} & {-19.71$^{+0.07}_{-0.09}$}& {7.89$^{+0.18}_{-0.45}$} & {$<$165.6}\\
{GN7\_14602} & {189.168625} & { 62.214306} & { 26.95$^{+0.25}_{-0.20}$} & {-20.05$^{+0.52}_{-0.27}$}& {7.91$^{+0.55}_{-0.52}$} & {$<$186.5}\\
{ERSPRIME\_5454792} & { 53.072163} & {-27.700729} & { 26.64$^{+0.14}_{-0.12}$} & {-20.70$^{+0.09}_{-0.09}$}& {7.92$^{+0.66}_{-0.69}$} & {$<$69.5}\\
{GN3\_5454841} & {189.214523} & { 62.305443} & { 27.62$^{+0.12}_{-0.11}$} & {-19.50$^{+0.08}_{-0.11}$}& {7.95$^{+0.35}_{-0.38}$} & {$<$240.2}\\
{GS4\_26328} & { 53.164701} & {-27.788230} & { 27.43$^{+0.04}_{-0.04}$} & {-19.80$^{+0.04}_{-0.01}$}& {7.95$^{+0.17}_{-0.16}$} & {$<$73.6}\\
{GN3\_34236} & {189.287003} & { 62.318020} & { 26.49$^{+0.19}_{-0.16}$} & {-20.71$^{+0.12}_{-0.13}$}& {8.00$^{+0.38}_{-0.41}$} & {$<$62.4}\\
{GN4\_5439303} & {189.333542} & { 62.254501} & { 28.57$^{+0.20}_{-0.17}$} & {-18.75$^{+0.16}_{-0.18}$}& {8.15$^{+0.61}_{-0.97}$} & {$<$516.4}\\
{GN3\_33315} & {189.224442} & { 62.311329} & { 25.94$^{+0.02}_{-0.02}$} & {-21.39$^{+0.02}_{-0.02}$}& {8.18$^{+0.10}_{-0.09}$} & {$<$38.7}\\
\enddata
\tablenotetext{}{
$^{a}$\footnotesize The listed IDs are encoded with their observed CLEAR fields and matched 3D-HST IDs. We use 5400000 + IDs from \cite{Finkelstein2015a} for 3D-HST unmatched objects.\\
$^{b}$$M_{\text{UV}}$ is estimated from the averaged flux over a 1450 -- 1550\AA\ bandpass from the best-fit galaxy SED model.\\
$^{c}$We present the 1$\sigma$ range of $z_{\text{phot}}$.\\
$^{d}$$3\sigma$ upper limits, measured from the median flux limits from individual spectra. \\
$^{e}$This object is not included in the analysis for the Ly$\alpha$ EW distribution and the IGM transmission due to the tentative emission-line detection.
}
\end{deluxetable*}

\bibliographystyle{aasjournal}
\bibliography{references}
\end{document}